\documentclass[lettersize,journal]{IEEEtran}
\usepackage[utf8]{inputenc}
\usepackage[sort&compress,numbers]{natbib}
\usepackage{babel}
\usepackage{acronym}
\usepackage{float}
\usepackage{placeins}
\usepackage[noend]{algpseudocode}
\usepackage{amsmath,amssymb,amsfonts}
\usepackage{amsmath}
\usepackage{mathtools}

\usepackage{scalerel,stackengine}
\stackMath
\newcommand\reallywidehat[1]{%
\savestack{\tmpbox}{\stretchto{%
  \scaleto{%
    \scalerel*[\widthof{\ensuremath{#1}}]{\kern-.6pt\bigwedge\kern-.6pt}%
    {\rule[-\textheight/2]{1ex}{\textheight}}%
  }{\textheight}%
}{0.5ex}}%
\stackon[1pt]{#1}{\tmpbox}%
}
 \usepackage{nccmath}
\usepackage{graphicx}
\usepackage{textcomp}
\usepackage{xcolor}
\usepackage{algpseudocode}
\usepackage{cleveref}

\usepackage{algorithm}
\usepackage{placeins}
\usepackage[noend]{algpseudocode}
\usepackage{amsmath,amssymb,amsfonts}
\usepackage{amssymb}%
\usepackage{pifont}%
\newcommand{\bluecheck}{\ding{51}}%
\newcommand{\xmark}{\ding{55}}%
\usepackage{mathtools}
\usepackage{textcomp}
\usepackage{xcolor}
\usepackage{algpseudocode}
\makeatletter
\def\BState{\State\hskip-\ALG@thistlm}
\makeatother

\DeclareMathOperator{\Exp}{\mathbb{E}}

\DeclareMathAlphabet{\mathbit}{OML}{cmr}{bx}{it}
\crefname{figure}{Fig.}{Figs.}%
\newcommand{\B}[1]{\mathbf{#1}}

\acrodef{MRT}{maximum ratio transmitter}
\acrodef{D2D}{device-to-device}
\acrodef{UAV}[UAV]{unmanned aerial vehicle}
\acrodef{CR}{cognitive radio}
\acrodef{CSI}{channel state information}
\acrodef{ICSI}{imperfect channel state information}
\acrodef{UPA}{uniform planar array}
\acrodef{ISI}{inter-symbol interference}
\acrodef{OFDM}{orthogonal frequency division multiplexing}
\acrodef{QoS}{quality of service}
\acrodef{MISO}{multiple-input single-output}
\acrodef{SIMO}{single-input multiple-output}
\acrodef{AoA}{angles of arrival}
\acrodef{AoD}{angles of departure}
\acrodef{PG}{projected gradient}
\acrodef{SNR}{signal-to-noise ratio}
\acrodef{BS}{base station}
\acrodef{MSE}{mean square error}
\acrodef{MAC}{multiple access channel}
\acrodef{MIMO}{multiple-input multiple-output}
\acrodef{SISO}{single-input single-output}
\acrodef{MMSE}{minimum mean square error}
\acrodef{LS}{least squares}
\acrodef{AWGN}{additive white Gaussian noise}
\acrodef{NOMA}{non-orthogonal multiple access}
\acrodef{IRS}{intelligent reflecting surface}
\acrodef{mmWave}{millimeter-wave}
\acrodef{SVD}{singular value decomposition}
\acrodef{BC}{broadcast}
\acrodef{MU}{multiuser}
\acrodef{DPC}{dirty paper coding}
\acrodef{AF O-Ps}{Amplify-and-Forward with optimized precoders}
\acrodef{R-IRS O-Ps}{Random IRS with optimized precoders}
\acrodef{O-IRS MRT-Ps}{Optimized IRS with MRT precoders}
\acrodef{No-IRS MRT-Ps}{No IRS with MRT precoders}

\usepackage{tikz}
\usetikzlibrary{calc}

\newcommand{\positiontextbox}[4][]{%
  \begin{tikzpicture}[remember picture,overlay]
    \node[inner sep=3pt, fill=yellow,align=left,draw,line width=1pt,#1] at ($(current page.north west) + (#2,-#3)$) {\parbox{.95\paperwidth}{#4}};
  \end{tikzpicture}%
}

\begin{document}

\onecolumn
\begingroup

\setlength\parindent{0pt}
\fontsize{14}{14}\selectfont

\vspace{1cm} 
\textbf{This is an ACCEPTED VERSION of the following published document:}

\vspace{1cm} 
D. Pérez-Adán, M. Joham, Ó. Fresnedo, J. P. González-Coma, L. Castedo and W. Utschick, ``Alternating Minimization for Wideband Multiuser IRS-Aided MIMO Systems Under
Imperfect CSI'', \textit{IEEE Transactions on Signal Processing}, vol. 72, pp. 99-114, 2024, doi:
10.1109/TSP.2023.3336166.

\vspace{1cm} 
Link to published version: https://doi.org/10.1109/TSP.2023.3336166

\vspace{3cm} 

\textbf{General rights:}

\vspace{1cm} 
\textcopyright 2023 IEEE. This version of the article has been accepted for publication, after peer review. Personal use of this material is permitted. Permission from IEEE must be obtained for all other uses, in any current or future media, including reprinting/republishing this material for advertising or promotional purposes, creating new collective works, for resale or redistribution to servers or lists, or reuse of any copyrighted component of this work in other works.
\twocolumn
\endgroup
\clearpage

\title{Alternating Minimization for Wideband Multiuser IRS-aided MIMO Systems under Imperfect CSI}

\author{Darian Pérez-Adán,~\IEEEmembership{Member,~IEEE,} Michael Joham,~\IEEEmembership{Member,~IEEE,} Óscar Fresnedo,~\IEEEmembership{Member,~IEEE,} José P. González-Coma, Luis Castedo,~\IEEEmembership{Senior Member,~IEEE} and Wolfgang Utschick,~\IEEEmembership{Fellow Member,~IEEE}
\thanks{Darian Pérez-Adán, Óscar Fresnedo and Luis Castedo are with the Department of Computer Engineering, University of A Coruña, CITIC, Spain, e-mail: \{d.adan, oscar.fresnedo,  luis\}@udc.es. José P. González-Coma is with the Defense University Center at the Spanish
Naval Academy. jose.gcoma@cud.uvigo.es. Michael Joham and Wolfgang Utschick are with the Department of Electrical and Computer Engineering, Technical University of Munich, Munich, Germany. \{joham, utschick\}@tum.de}%
}

\markboth{Journal of \LaTeX\ Class Files,~Vol.~14, No.~8, August~2021}%
{Shell \MakeLowercase{\textit{et al.}}: Alternating Minimization for Wideband Multiuser IRS-aided MIMO Systems under Imperfect CSI}

\maketitle

\positiontextbox{10.75cm}{27cm}{\footnotesize \textcopyright 2023 IEEE. This version of the article has been accepted for publication, after peer review. Personal use of this material is permitted. Permission from IEEE must be obtained for all other uses, in any current or future media, including reprinting/republishing this material for advertising or promotional purposes, creating new collective works, for resale or redistribution to servers or lists, or reuse of any copyrighted component of this work in other works. Published version:
https://doi.org/10.1109/TSP.2023.3336166}

\begin{abstract}
This work focuses on wideband \ac{IRS}-aided multiuser MIMO systems. One of the major challenges of this scenario is the joint design of the frequency-dependent \ac{BS} precoder and user filters, and the \ac{IRS} phase-shift matrix which is frequency flat and common to all the users. 
In addition, we consider that the \ac{CSI} is imperfect at both the transmitter and the receivers. A statistical model for the imperfect \ac{CSI} is developed and exploited for the system design.
A \ac{MMSE} approach is followed to determine the \ac{IRS} phase-shift matrix, the transmit precoders, and the receiving filters. The \ac{BC}-\ac{MAC} duality is used to solve the optimization problem following an alternating minimization approach. Numerical results show that the proposed approach leads to substantial performance gains with respect to baseline strategies that neglect the inter-user interference and do not optimize the \ac{IRS} phase-shift matrix. Further performance gains are obtained when incorporating into the system design the statistical information of the channel estimation errors.
\end{abstract}


\section{Introduction}
\IEEEPARstart{A}{n} 
\acf{IRS} is a metasurface comprising a multitude of low-cost passive reflective elements whose response can be configured in real-time \cite{8917871,9326394,8936989,basar2019wireless,gong2020toward,9205201}. 
\acp{IRS} are attracting  significant attention as a key enabling technology for \ac{MIMO} systems to reach the capacity requirements demanded by the next generations of wireless communication systems \cite{6736746,rappaport_millimeter_2014, heath_overview_2016,6732923,6824752}. 

The \ac{IRS} technology offers tremendous benefits for various application scenarios \cite{perez2021intelligent}. In \ac{D2D} networks \cite{mustafa_separation_2016}, \acp{IRS} can be utilized to cancel interference, support low-power transmission, and enhance individual data links. In \ac{CR} networks, the \acp{IRS} play a crucial role in increasing the degrees of freedom and improving the efficiency of secondary transmissions \cite{yuan_intelligent_2021,allu2023robust}. Another interesting application of \acp{IRS} is in cellular network systems with cell edge users. These users often suffer from high signal attenuation from the \ac{BS} and co-channel interference from nearby \acp{BS}. By deploying \acp{IRS} in such scenarios, the coverage area in the cellular network can be expanded as \acp{IRS} efficiently reflect the signal from the BS to the cell edge user, compensating for signal attenuation and interference, as demonstrated in \cite{ding_simple_2020}. Moreover, \acp{IRS} find intriguing applications in \ac{UAV} networks. Integrating \acp{IRS} in \ac{UAV} networks can significantly enhance communication quality between \acp{UAV} and ground users. This enhancement becomes crucial for optimizing both UAV trajectories and overall system performance, as noted in \cite{guo_learning-based_2021}.

Different \ac{IRS}-aided wireless communication systems have been considered in the literature. In \cite{bjornson2019intelligent}, a single-user \ac{SISO} system with a direct channel between both communication ends is considered. In this specific case, the optimal \ac{IRS} phase-shift matrix is the one that aligns the reflected rays to the direct path between the transmitter and the receiver. However, this solution is not applicable to \ac{MU} \ac{MIMO} systems where a common \ac{IRS} response must be designed for all users. 

In \cite{ozdogan2020using}, the rank improvement of a downlink \ac{IRS}-aided \ac{MIMO} system is exploited to obtain capacity gains in a single-user scenario. The authors in \cite{jung2020asymptotic} have studied the asymptotic achievable rate of the downlink of an \ac{IRS}-aided \ac{MU} \ac{MISO} system where some users are supported by \acp{IRS} while others directly communicate to the \ac{BS}. The authors propose a modulation scheme that results in achievable sum rates larger than those obtained in non-\ac{IRS}-aided schemes.
In \cite{9483903}, the authors investigate \acs{MU} \acs{MISO} downlink communications assisted by a self-sustainable \ac{IRS}. The reflecting elements of the \ac{IRS} are classified into two categories: energy harvesting elements and communication elements. As a consequence, the \ac{IRS} is capable of both reflecting signals and harvesting energy from the received signals. The primary objective of  \cite{9483903} is to maximize the system rate, and to achieve this goal, the authors propose an iterative algorithm that provides a suboptimal solution to the design problem.
In \cite{fu2019intelligent}, an \ac{IRS}-aided \ac{MU} \ac{NOMA} \ac{MISO} downlink system\textemdash which again enables communication over both the \ac{IRS}-aided and the direct channels\textemdash is addressed. The approach considers the joint optimization of the \ac{BS} precoders and the \ac{IRS} phase-shift matrix to minimize the total transmission power. The authors in \cite{yan2020passive} consider an \ac{IRS}-aided \ac{MU} \ac{MISO} system %
with an on/off modulation at the reflective elements. 
In \cite{10054092}, the authors focus on the joint optimization of the IRS phase-shift matrix and the MIMO precoders in an IRS-aided \ac{MU} MIMO system. Their objective is to maximize the system sum-rate for each channel realization using reinforcement learning strategies. The proposed algorithms are particularly suitable for scenarios with high \ac{MU} interference.
Nevertheless, there are \ac{IRS}-aided wireless communication use cases that still remain unexplored. For instance, none of the works in \cite{bjornson2019intelligent,ozdogan2020using,jung2020asymptotic,fu2019intelligent,yan2020passive,9483903} consider \ac{MU} \ac{MIMO} setups, only \cite{10054092} considers this scheme. They also do not consider wideband transmissions, a relevant feature when communicating through \ac{mmWave} bands. 

Only a few papers have explored the use of IRS-aided systems in wideband scenarios. An example is \cite{yue2023ris} where an uplink \ac{MU} \ac{SIMO} \ac{OFDM} scheme is considered. The focus in \cite{yue2023ris} is on minimizing the total transmit power by jointly designing the precoders and optimizing the passive beamforming carried out by the \ac{IRS}.

Wideband systems offer several advantages over conventional narrowband systems, including higher data rates, improved spectral efficiency, and better interference management. Their ability to operate over wider frequency ranges and efficiently utilize wide bandwidths makes them well-suited for modern high-speed communication applications, particularly in scenarios with large data rate demands and multiple users  \cite{schulze2005theory}.

Employing \acp{IRS} in wideband \ac{MU} systems provides several benefits that can significantly enhance the overall system performance. \acp{IRS} can effectively control the propagation environment by intelligently reflecting and redirecting incident signals. This leads to improved signal strength at the receivers, mitigating path loss and signal fading, which is particularly beneficial in wideband scenarios where frequency-selective fading may occur. Additionally, \acp{IRS} can be utilized to suppress unwanted interference among users, especially in densely populated \ac{MU} scenarios. By adjusting the phase shifts of the reflecting elements, the system can actively control signal directions and reduce inter-user interference \cite{9205201}.

In this case of wideband \ac{MU} systems, the design of \ac{IRS}-aided wireless systems is challenging because the \ac{IRS} phase shift matrix is frequency flat while the wireless channels are frequency selective. 

Regarding \ac{CSI}, few works in the literature consider the impact of channel estimation errors when designing robust IRS-aided systems. For instance, in  \cite{9117093}, the authors investigate \ac{MMSE}-optimal beamforming in a narrowband single-user \ac{MISO} system while taking into account imperfect \ac{CSI}. Another work, \cite{9110587}, focuses on the design of an \ac{IRS}-aided narrowband \ac{MU} \ac{MISO} system and considers the effect of imperfect \ac{CSI} in their analysis. 
The authors in \cite{9348255} propose a robust \ac{IRS} design (with on/off reflection) for a narrowband single-user \ac{MISO} system.  They use the penalized Dinkelbach method to optimize the IRS reflection coefficients to maximize the achievable rate under \ac{CSI} errors. 
In \cite{9180053}, the authors address a narrowband MU MISO scheme under imperfect CSI, considering only single-antenna users and the error model associated with the cascaded channels.
In \cite{9483903}, the authors also consider a robust design of a narrowband \ac{MU} \acs{MISO} system based on a suboptimal solution to a rate maximization problem. In
\cite{9374975}, the authors consider a narrowband MU (single-antenna) MISO IRS-aided system with on/off reflections at the IRS and focus on maximizing the achievable rate under imperfect CSI.
In \cite{allu2023robust}, the authors investigate a \ac{MIMO} \ac{CR} system, where a secondary transmitter serves multiple secondary receivers concurrently. The design of the IRS matrix, precoding, and reception filters is considered under the presence of imperfect CSI at the secondary transmitter by using a norm-bounded error model.
In \cite{9851661}, the authors address a narrowband vehicular MU IRS-aided communication system under imperfect CSI. They approach the joint precoder and IRS design by considering the outage probability constraint. Additionally, this work explores the concept of an active IRS, simplifying the design process under imperfect CSI by estimating the cascaded channel in two separate steps as in \cite{9110587}: first, estimating the BS-IRS channel, and then estimating the IRS-user channels. However, this channel estimation strategy is impractical for passive IRSs.
To highlight the novelty of the proposed scheme, a brief comparison with existing works in the literature is provided in  \Cref{Tablanovedad}.

\begin{table*}[htpb]
\vspace{-2mm}
	\centering\caption{\label{Tablanovedad} Contrasting the  contributions of our scheme to the state-of-the-art.}
\centering
\vspace{-2mm}
\def\arraystretch{1.35}
\begin{tabular}{c c c c c c c c| c c c c c c c c c c|} \cline{2-17} 
\multicolumn{1}{c}{} & \multicolumn{7}{c}{\textbf{Perfect CSI}} & \multicolumn{9}{|c}{\textbf{Imperfect CSI}}  \\  \cline{2-17} 
                  &\cite{bjornson2019intelligent} & \cite{ozdogan2020using} & \cite{jung2020asymptotic} &   \cite{fu2019intelligent} & \cite{yan2020passive} &   \cite{10054092}& \cite{yue2023ris} & \cite{allu2023robust}     & \cite{9483903} & \cite{9117093} & \cite{9110587} &\cite{9348255} &\cite{9180053} & \cite{9374975}   & \cite{9851661} &Our scheme \\ \hline
           MIMO            & \xmark & \bluecheck   & \xmark   &\xmark  & \xmark  &  \bluecheck &  \xmark&  \bluecheck & \xmark& \xmark &   \xmark &    \xmark &   \xmark & \xmark&  \xmark& \bluecheck \\ 
          Multiuser       &\xmark      & \xmark   & \bluecheck  & \bluecheck & \bluecheck  & \bluecheck  & \bluecheck &  \xmark &\bluecheck &\xmark &     \bluecheck  &  \xmark &\bluecheck & \bluecheck&  \bluecheck& \bluecheck \\ 
           Wideband           &\xmark &   \xmark   & \xmark  &   \xmark  &  \xmark &  \xmark & \bluecheck & \xmark  &\xmark & \xmark & \xmark & \xmark & \xmark  &\xmark& \xmark&  \bluecheck \\ 
           mmWave       & \xmark      & \xmark       & \xmark & \xmark & \xmark & \xmark  & \xmark &  \xmark & \bluecheck& \xmark & \xmark   & \xmark & \xmark &\xmark & \xmark & \bluecheck \\

           Passive IRS         &\bluecheck   &  \bluecheck &   \bluecheck        & \bluecheck      &\bluecheck  & \bluecheck & \bluecheck & \xmark& \bluecheck &\bluecheck  & \bluecheck   &\bluecheck  &  \bluecheck &\bluecheck&  \xmark & \bluecheck \\ \hline
\end{tabular}
\end{table*}

\subsection{Contributions and organization}
Unlike the aforementioned references \cite{allu2023robust,bjornson2019intelligent,ozdogan2020using,jung2020asymptotic,fu2019intelligent,yan2020passive,9483903,yue2023ris,10054092, 9117093,9110587,9180053,9348255,9374975,9851661}, this work considers an IRS-aided wideband MU MIMO system with a passive IRS and imperfect CSI. We address the joint optimization of the frequency-dependent precoders/filters and the frequency-flat IRS phase-shift matrix, assuming imperfect CSI at the transmitter and receivers.
We consider the intrinsic characteristics of IRS-aided systems, especially those with passive IRSs, to model imperfect CSI and develop a robust solution. In our approach, the estimation errors jointly affect both links in the cascaded channel, since we can only obtain an aggregated CSI from the IRS-involved channels, unlike the approaches in
\cite{9851661,9110587}.

We develop an alternating approach that minimizes the \ac{MSE} between the original and estimated symbols by exploiting the \ac{BC}-\ac{MAC} duality. At each iteration, the updates are based on the \ac{BC}-\ac{MAC} duality and a \ac{PG} algorithm.
We consider the specific characteristics of the \ac{MAC}-\ac{BC} and \ac{BC}-\ac{MAC} dualities when considering the deployment of an \ac{IRS} and the conformation of the cascaded channel.
The proposed alternating minimization algorithm iterates over the precoders/combiners in both the downlink and the uplink systems at each subcarrier, while also performing iterations over the frequency-flat IRS using a gradient descent step.

The proposed approach leads to better performance in terms of achievable sum-rate and \ac{MSE} over baseline strategies. More specifically, the main contributions of this work are the following:

\begin{itemize}
    \item 
{{We jointly design frequency-dependent precoders/filters and the frequency-flat IRS phase-shift matrix in a multistream wideband IRS-aided MU MIMO downlink by exploiting BC-MAC duality for aggregated imperfect CSI in passive IRS-aided systems. We also employ a PG algorithm to configure the IRS phase-shift matrix.}}
 \item 
{{We have developed an alternating minimization method to design the wideband precoders/filters and the frequency-flat IRS phase-shift matrix under imperfect CSI in the IRS-involved channels. The key aspect of this method is its consideration of the statistics of channel estimation errors to improve system performance. Moreover, this approach is independent of the specific statistics of the channel estimation errors.}}
\end{itemize}

The remainder of this work is organized as follows. The system model is described in \Cref{Section II} while the \ac{MMSE} design is explained in \Cref{Section III}. An alternating minimization algorithm for the joint computation of the wideband \ac{BS} precoders and user combiners, as well as the frequency-flat \ac{IRS}-phase shift matrix, is described in \Cref{Section IV} by considering imperfect \ac{CSI}. Convergence analysis for the proposed  algorithm is carried out in \Cref{Section_conv}. Computational complexity analysis is exposed in \Cref{Section V}. Simulation results are presented in \Cref{Section VI} and, finally, \Cref{Section VII} is devoted to the conclusions.

\subsection{Notation}
The following notation will be employed throughout the entire work: $a$ is a scalar, $\mathbf{a}$ is a vector, and $\mathbf{A}$ is a matrix. $[\mathbf{{A}}]_{i,j}$ is the entry on the $i$-th row and the $j$-th column of $\mathbf{A}$.
Transpose and conjugate transpose of $\mathbf{A}$ are represented by $\mathbf{A}^{\operatorname{T}}$ and $\mathbf{A}^{*}$, respectively. ${\|\mathbf{A} \|}_{\operatorname{F}}^2$ and $\text{tr}[\mathbf{A}]$ denote the Frobenius norm and the trace of $\mathbf{A}$, respectively. $\mathbf{A}^{\dagger}$ represents the pseudoinverse of $\mathbf{A}$. Calligraphic
letters are employed to denote sets.
The operator
$\operatorname{blkdiag\;(\cdot)}$ constructs a block diagonal matrix from its input matrices. Finally, the expectation is denoted by $\Exp [\cdot]$, $\circledast$ is the column-wise Khatri-Rao product, and $\otimes$ represents the Kronecker product.

\begin{figure}[h!]
\centering
\includegraphics[width=0.99\linewidth]{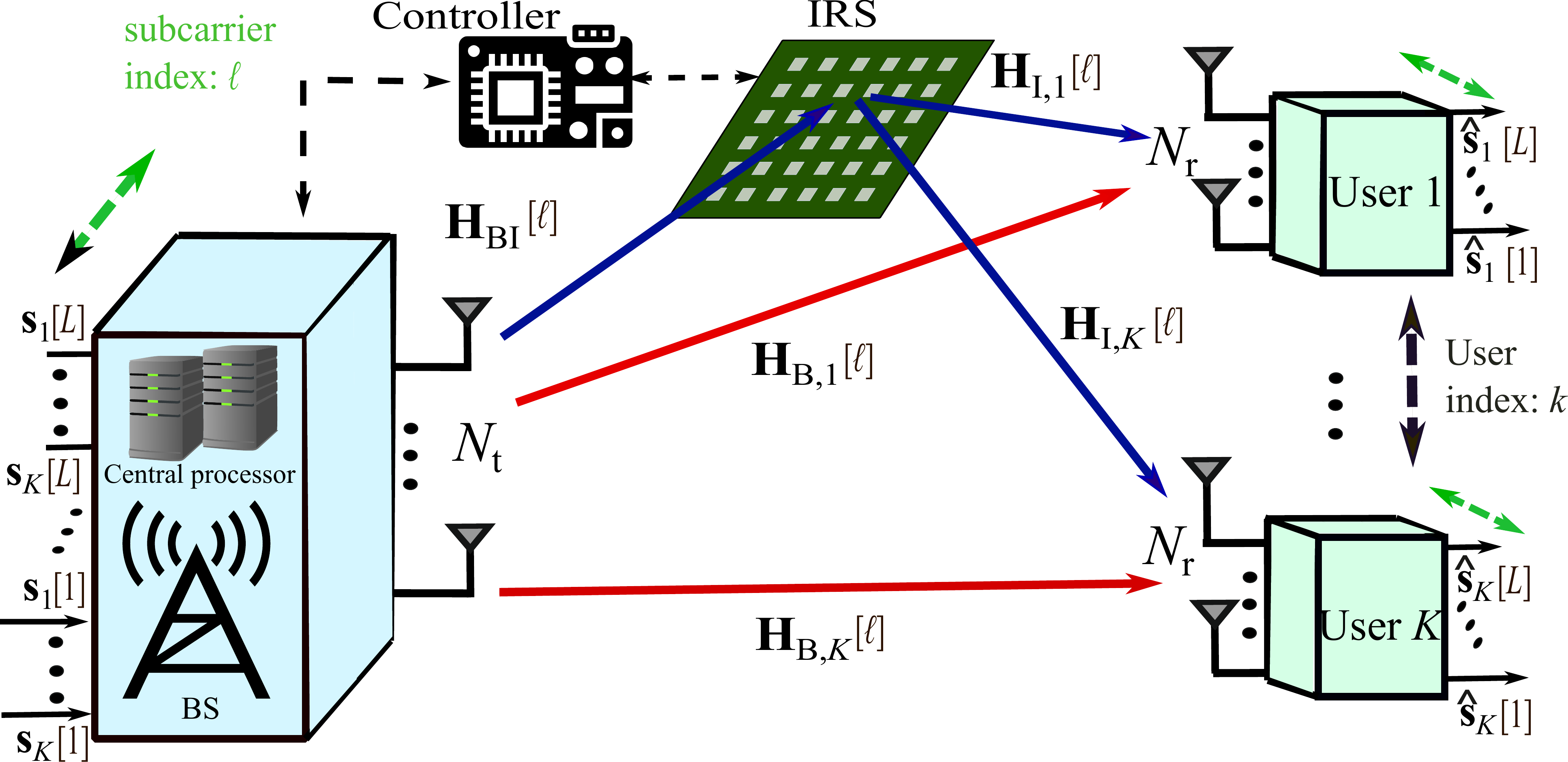}
\vspace{-3mm}
\caption{Block diagram of the multistream wideband downlink \ac{MU} \ac{IRS}-aided \ac{mmWave} \ac{MIMO} system.}\label{Fig01}
\end{figure}

\section{System Model}\label{Section II}
Let us consider the \ac{IRS}-aided \ac{MU} \ac{MIMO} downlink shown in \Cref{Fig01} where a common \ac{BS} with $N_{\text{t}}$ antennas communicates  with $K$ users with $N_{\text{r}}$ antennas each. The \ac{BS} sends wideband \ac{OFDM} symbols to fully exploit the large bandwidths available in \ac{mmWave}. The wireless channels between the \ac{BS}, the \acp{IRS} and the users are assumed to be frequency selective. The \ac{OFDM} modulation is also assumed to have $L$ subcarriers and a cyclic prefix long enough to avoid \ac{ISI}. This way, the frequency-selective channels are decomposed into $L$ parallel narrowband subchannels, each experiencing a different frequency response.

We also consider multistream transmission in such a way that the \ac{BS} allocates ${N_{\text{s},k}}[\ell]$ streams to be transmitted to the user $k \in \{1, \ldots ,K\}$ at the subcarrier $\ell \in \{1,\ldots,L\}$. The total number of streams allocated at subcarrier $\ell$ is $N_{\text{s}}[\ell]=\sum_{k=1}^K {N_{\text{s},k}}[\ell]$ and the total number of streams considering all the subcarriers is  $N_{\text{s}}=\sum_{\ell=1}^L {N_{\text{s}}}[\ell]$. At each channel use, the \ac{BS} transmits to the $k$-th user at subcarrier $\ell$ the vector of zero-mean symbols $\mathbf{s}_k^{}[\ell]={\left[{s_k}_{1}[\ell],{s_k}_{2}[\ell],\ldots,{s_k}_{N_{\text{s},k}[\ell]}[\ell]\right]}^{\operatorname{T}}$ with  $\mathbb{E}\left[\mathbf{s}_k[\ell]\mathbf{s}^*_k[\ell]\right]=\mathbf{I}_{N_{\text{s},k}[\ell]}$. We assume there is no blockage between the \ac{BS} and the users, i.e., a direct channel is available between the \ac{BS} and the $k$-th user whose response at subcarrier $\ell$ is ${{\mathbf{{H}}^{}_{\text{B},{k}}}}[\ell]\in \mathbb{C}^{N_{\text{r}}\times N_{\text{t}}}$. 

In addition, the \ac{IRS} introduces an additional cascaded \ac{BS}-\ac{IRS}-User link. The channel responses from the \ac{IRS} to the $k$-th user and from the \ac{BS} to the \ac{IRS} at subcarrier $\ell$ are represented by ${{\mathbf{{H}}^{}_{\text{I},{k}}}}[\ell]\in \mathbb{C}^{N_{\text{r}}\times N_{\text{}}}$ and ${{\mathbf{{H}}^{}_{\text{BI}}}_{}}[\ell]\in \mathbb{C}^{N_{\text{}}\times N_{\text{t}}}$, respectively. The deployed \ac{IRS} is assumed to have $N$ passive elements and its phase-shift matrix is represented by the diagonal matrix $\mathbf{\Theta}^{}=\text{diag}(\boldsymbol{\nu})\in\mathcal{D}$, where $\boldsymbol{\nu}=[\nu_1,\ldots,\nu_N]^{\operatorname{T}}=[e^{\operatorname{j}\theta_1},\ldots, e^{\operatorname{j}\theta_{N}}]^{\operatorname{T}}$ and $\theta_{n}\in[0,2\pi)\;\forall n$ is the phase shift introduced by the $n$-th \ac{IRS} element. $\mathcal{D}\in \mathbb{C}^{N\times N}$ is the set of feasible \ac{IRS} matrices, i.e., the set of diagonal matrices with unit magnitude diagonal entries. The cascaded \ac{BS}-\ac{IRS}-user channel response at subcarrier $\ell$ is given by ${\mathbf{{H}}_{\text{c},{k}}}[\ell]={\mathbf{{H}}_{\text{I},{k}}}[\ell]\mathbf{\Theta}^{}{{\mathbf{{H}}^{}_{\text{BI}}}}[\ell]=\sum^{N_{}}_{n=1}{\nu}_n{\mathbf{{h}}_{\text{I},{k,n}}}[\ell]{\mathbf{{h}}^{\operatorname{T}}_{\text{BI},{n}}}[\ell]$, where $\mathbf{h}_{\text{I},k,n}[\ell]$ and $\mathbf{h}_{\text{BI},n}[\ell]$ denote the $n$-th column of $\mathbf{H}_{\text{I},k}[\ell]$ and $\mathbf{H}^{\operatorname{T}}_{\text{BI}}[\ell]$, respectively. We highlight that the frequency response of the \ac{IRS} is nominally flat and, thus, common to all the subcarriers and users. As explained later, this circumstance makes the design of the \ac{IRS}-aided communication system significantly more difficult.

The \ac{BS} employs the  linear precoder $\mathbf{P}_k[\ell]\in \mathbb{C}^{N_{\text{t}}\times N_{{\text{s},k}}[\ell]}$ to communicate with the $k$-th user at subcarrier $\ell$. 
As in \cite{8323164,9374961}, these precoders are subject to the per-subcarrier transmission power constraint $\sum_{k=1}^{K} \parallel{\mathbf{P}_{k}[\ell]}\parallel^2_{\operatorname{F}}\leq P_{\text{T}}[\ell]$ where $P_{\text{T}}[\ell]$ is the available power at subcarrier $\ell$. For the sake of simplicity, and due to the optimality at high \acp{SNR}, we will assume that $P_{\text{T}}[\ell]=\frac{P_{\text{T}}}{L}$, where $P_{\text{T}}$ is the total power available at transmission.\footnote{We assume that all the computation related to the system design is performed at the \ac{BS}, which serves as the resource allocator.} The signal received by the $k$-th user at subcarrier $\ell$ is given by 
\begin{equation}\label{1}
\mathbf{y}_k[\ell]=\left(\underbrace{{\mathbf{{H}}_{\text{B},{k}}}[\ell]}_{\textrm{Direct link}}+\underbrace{{\mathbf{{H}}_{\text{I},{k}}}[\ell]\mathbf{\Theta}^{}{{\mathbf{{H}}^{}_{\text{BI}}}}[\ell]}_{\textrm{\ac{IRS}-aided link}}\right)\sum_{u=1}^{K} {\mathbf{P}}_{u}[\ell]\mathbf{s}_u[\ell]+\boldsymbol{\eta}_k[\ell],
\end{equation}
where $\boldsymbol{{\eta}}_k[\ell]=\left[{\eta_k}_{1}, {\eta_k}_{2},\ldots,{\eta_k}_{N_{\text{r}}}\right]^{\operatorname{T}}$ represents the complex-valued \ac{AWGN} modeled as $\boldsymbol{\eta}_k[\ell]\sim \mathcal{N}_{\mathbb{C}} (\boldsymbol{0}, \mathbf{C}_{{{\eta}_k}})$.
Finally, the equivalent channel response corresponding to the $k$-th user at subcarrier $\ell$ is defined as
\begin{align}\label{mm}
\notag{\mathbf{H}_{\text{e},k}}[\ell]&={{\mathbf{H}_{\text{B},k}}}[\ell]+{{\mathbf{H}_{\text{I},k}}}[\ell]\mathbf{\Theta}^{}{\mathbf{H}^{}_{\text{BI}}}[\ell]\\
&={{\mathbf{H}_{\text{B},k}}}[\ell]+\sum^{N_{}}_{n=1}{\nu}_n\mathbf{{h}}_{\text{I},k,n}[\ell]\mathbf{{h}}^{\operatorname{T}}_{\text{BI},n}[\ell],
\end{align}
{where $n \in \{1,\ldots,N\}$ stands for the index of each \ac{IRS} element to modify each channel path by introducing the phase shift corresponding to the coefficient ${\nu}_n=e^{\operatorname{j} \theta_n}$.

\subsection{Channel estimation}\label{ches}
To carry out the \ac{CSI} estimation, pilot symbols without precoding are transmitted by the \ac{BS} during $N_{\text{p}}$ channel uses. The matrix $\mathbf{X}_{}[\ell]\in \mathbb{C}^{ N_{{\text{p}}}\times N_{\text{t}}}$ comprises all the pilots transmitted %
at subcarrier $\ell$. Each row vector in $\mathbf{X}_{}[\ell]$ represents the pilots transmitted in the corresponding  channel use.\footnote{The reduction of estimated overhead in \ac{MU} schemes is beyond the scope of the paper. However, interested readers can refer to \cite{9103231} for further exploration of this topic.} As discussed in \cite{joham2022estimation}, we assume that $N_{\text{p}}=N_{\text{t}}$ and $\mathbf{X}[\ell]$ is a weighted unitary matrix such that $\parallel \mathbf{X}[\ell]\parallel_{\operatorname{F}}^2=\frac{P_{\text{T}}}{L}N_{\text{p}}$.

The pilot symbols received by user $k$ at subcarrier $\ell$ are collected in the matrix $\mathbf{Y}^{\prime}_k[\ell]\in\mathbb{C} ^{N_{\text{r}}\times N_{\text{p}}}$ represented as
\begin{equation}\label{1ep}
\mathbf{Y}_k^{\prime}[\ell]=\left(\underbrace{{\mathbf{{H}}_{\text{B},k}}[\ell]}_{\textrm{Direct link}}+\underbrace{ {\mathbf{{H}}_{\text{I},{k}}}[\ell]\mathbf{\Theta}^{}{{\mathbf{{H}}^{}_{\text{BI}}}}[\ell]}_{\textrm{\ac{IRS}-aided link}}\right) {\mathbf{X}}[\ell]+\mathbf{{N}}_k^{\prime}[\ell].
\end{equation}
Columns of $\mathbf{{N}}_k^{\prime}[\ell]$ are mutually independent \ac{AWGN} vectors distributed as $\mathcal{N}_{\mathbb{C}}(\boldsymbol{0},\mathbf{C}_{\eta_k})$. By vectorizing $\mathbf{Y}^{\prime}_k[\ell]$ we get  (cf. \cite[Theorem 3.13]{1084534}) 
\begin{equation}\label{vecto}
    \mathbf{y}^{\prime}_k[\ell]=\left(\mathbf{X}[\ell]\otimes\mathbf{I}_{N_{\text{r}}}\right)\mathbf{H}_k[\ell]\boldsymbol{\nu}^{\prime}+\boldsymbol{\eta}^{\prime}_k[\ell],
\end{equation}
where $\boldsymbol{\nu}^{\prime}=\left[1,\boldsymbol{\nu}^{\operatorname{T}}\right]^{\operatorname{T}} \in\mathbb{C}^{N+1}$, 
\begin{equation}
\mathbf{H}_k[\ell]=\left[{\mathbf{h}_{\text{B},k}}[\ell],\mathbf{H}_{\text{BI}}^{\operatorname{T}}[\ell]\circledast {\mathbf{H}_{\text{I},k}}[\ell]\right]\in\mathbb{C} ^{N_{\text{r}}N_{\text{t}}\times N+1}\end{equation} 
stacks both the vectorized versions of the direct channel ${\mathbf{h}_{\text{B},k}}[\ell]=\text{vec}(\mathbf{H}_{\text{B},k}[\ell])$ and the cascaded channel, and $\boldsymbol{\eta}^{\prime}_k \sim \mathcal{N}_{\mathbb{C}}(0,\mathbf{I}_{N_{\text{t}}}\otimes {\mathbf{C}_{\eta_k}})$.
Assuming $N_{\nu}$ different phase allocations during the estimation process to modify the paths of the whole channel, we have
\begin{equation}
    \mathbf{V}=\left[\boldsymbol{\nu}^{\prime}_1,\ldots,\boldsymbol{\nu}^{\prime}_{N_{\nu}}\right] \in\mathbb{C} ^{N+1 \times N_{\nu}}.
\end{equation}
As discussed in \cite{joham2022estimation}, we set $N_{\nu}=N+1$ and $\mathbf{V}$ is a weighted unitary matrix with unit-magnitude entries, e.g., a DFT or Hadamard matrix, such that $\parallel\mathbf{V}\parallel^2_{\operatorname{F}}=(N+1)^2$.
We next assume that the same pilot matrix $\mathbf{X}[\ell]$ is transmitted over the $N_{\nu}$ different phase allocations and that the corresponding received symbols are stacked in the following matrix
\begin{equation}\label{vecto1}
    \mathbf{Y}^{}_k[\ell]=\left(\mathbf{X}[\ell]\otimes\mathbf{I}\right)\mathbf{H}_k[\ell]\mathbf{V}+\mathbf{N}^{}_k[\ell] \in \mathbb{C}^{N_{\text{r}}N_{\text{p}}\times N_{\nu}}.
\end{equation}
 Note now that the channel response $\mathbf{H}_k[\ell]$ for the $k$-th user can be estimated following a \ac{LS} approach. Indeed, the \ac{LS} estimation of the channel response is given by
\begin{equation}\label{lse}
    \mathbf{\hat{H}}_{\text{LS},k}[\ell]=\left( \mathbf{X}[\ell] \otimes\mathbf{I}\right)^{\dagger}\mathbf{Y}_k[\ell]\mathbf{V}^{\dagger} \in \mathbb{C}^{N_{\text{r}}N_{\text{t}}\times N+1},
\end{equation}
and the \ac{LS} estimation error is given by the following matrix
\begin{equation}
{\mathbf{N}_{\text{LS},k}}[\ell]=\left(\mathbf{X}[\ell] \otimes \mathbf{I}\right)^\dagger \mathbf{N}[\ell]\mathbf{V}^{\dagger}\in \mathbb{C}^{N_{\text{r}}N_{\text{t}}\times N+1},
\end{equation}
whose independent columns  have the  distribution $\mathcal{N}_{\mathbb{C}}(\boldsymbol{0}, {\mathbf{C}_{\text{LS},k}})$ with the following covariance matrix
\begin{equation}
    \label{CovLS}
    {\mathbf{C}_{\text{LS},k}}\hspace{-1mm}=\hspace{-1mm} \Big(\left(\mathbf{X}^*[\ell]\mathbf{X}[\ell]\right)^{-1}\otimes \mathbf{C}_{\eta_k}\Big)\text{tr}\Big(\left(\mathbf{V}\mathbf{V}^*\right)^{-1}\Big)\hspace{-1mm}\in\hspace{-1mm} \mathbb{C}^{N_{\text{r}}N_{\text{t}}\times N_{\text{r}}N_{\text{t}}}.
\end{equation}
where $\mathbf{C}_{\eta_k} \in \mathbb{C}^{N_{\text{r}}\times N_{\text{r}}}$ is the receiving noise covariance matrix. 

Leveraging the assumptions for $\mathbf{X}[\ell]$ and $\mathbf{V}$, that is, $\mathbf{X}^*[\ell]\mathbf{X}[\ell]=\frac{P_{\text{T}}}{L}\mathbf{I}_{N_{\text{t}}}$ and $\mathbf{V}\mathbf{V}^*=(N+1)\mathbf{I}_{N+1}$, leads to 

$$
\mathbf{C}_{\text{LS},k}=\left( \frac{L}{P_{\text{T}}}\mathbf{I}_{N_{\text{t}}}\otimes \mathbf{C}_{\eta_k} \right),
$$
and every column of $\mathbf{N}_{\text{LS},k}[\ell]$ has the distribution $\mathcal{N}_{\mathbb{C}}(\boldsymbol{0}, \frac{L}{P_{\text{T}}}\mathbf{I}_{N_{\text{t}}}\otimes \mathbf{C}_{\eta_k})$.

According to the above analysis, the channel uncertainty can be modeled as an statistical error.
Therefore, channel realizations in the downlink can be decomposed as follows
$${{\mathbf{{H}}_{\text{B},k}}[\ell] =\mathbf{\hat{H}}_{\text{B},k}}[\ell] +\mathbf{E}_{\text{B},k}[\ell]$$ for the direct channels and $${\mathbf{{H}}_{\text{I},k}}[\ell]\mathbf{\Theta}^{}{{\mathbf{{H}}^{}_{\text{BI}}}}[\ell]=\sum^{N_{}}_{n=1}{\nu}_n\left({{\mathbf{\hat{H}}}_{\text{c},{k,{n}}}}[\ell]+{\mathbf{E}_{\text{c},k,n}}[\ell]\right),$$ for the cascaded channels
such that
\begin{equation}\label{llv}
   {{\mathbf{\hat{H}}}_{\text{c},{k,{n}}}}[\ell]= \reallywidehat{\mathbf{{h}}_{\text{I},k,n}[\ell]{\mathbf{{h}}^{\operatorname{T}}_{\text{BI},n}}[\ell]},\forall n=1,\ldots,N.
\end{equation}
Note that the estimate $\reallywidehat{\mathbf{{h}}_{\text{I},k,n}[\ell]{\mathbf{{h}}^{\operatorname{T}}_{\text{BI},n}}[\ell]}$ eventually is not rank-one due to the noise.
 Here,  $\mathbf{H}_{\text{I},k}[\ell]$, ${\mathbf{{H}}_{\text{BI}}}[\ell]$ and 
$\mathbf{H}_{\text{B},k}[\ell]$ are the true channels at subcarrier $\ell$ of user $k$
whereas $\mathbf{\hat{H}}_{\text{c},{k}}[\ell]$, and $\mathbf{\hat{H}}_{\text{B},k}[\ell]$ stand for the estimated channels at subcarrier $\ell$ of user $k$.

Recall that the estimations of the channels $\mathbf{{H}}_{\text{I},k}[\ell]$ and ${\mathbf{{H}}_{\text{BI}}}[\ell]$ have to be performed jointly due to the passive nature of the \ac{IRS}. The entries of $\mathbf{{E}}_{\text{B},k}[\ell]$ and ${\mathbf{{E}}_{\text{c},{k},n}}[\ell]$ are the estimation errors of the direct and cascaded channels, respectively. These errors are zero-mean Gaussian distributed with a covariance matrix $\mathbb{E}[{\mathbf{{e}}_{\text{B},k}}[\ell]{\mathbf{{e}}^*_{\text{B},k}}[\ell]]=\mathbf{C}_{\text{LS},k}, \forall \ell$ and $\mathbb{E}[{\mathbf{{e}}_{\text{c},k,n}}[\ell]{\mathbf{{e}}^*_{\text{c},k,n}}[\ell]]=\mathbf{C}_{\text{LS},k}, \forall \ell, n$, respectively, where ${\mathbf{{e}}_{\text{B},k}}[\ell]$ and ${\mathbf{{e}}_{\text{c},k,n}}[\ell]$ are the vectorized versions of ${\mathbf{{E}}_{\text{B},k}}[\ell],\forall\; k,\ell$ and ${\mathbf{{E}}_{\text{c},k,n}}[\ell],\forall\; k,n,\ell$, respectively. They correspond to the first and the $(n+1)$-th column of ${\mathbf{N}_{\text{LS},k}}[\ell]$, respectively. Therefore,
note that the mutually independent columns of $\mathbf{E}_{\text{B},k}[\ell]$ and $\mathbf{E}_{\text{c},k,n}[\ell]$ are distributed as $\mathcal{N}_{\mathbb{C}}(\mathbf{0},\frac{L}{P_{\text{T}}}\mathbf{C}_{\eta_k})$.
According to this error model, the next equivalence can be stated
\begin{equation}\label{heqv}
    {\mathbf{H}_{\text{e},k}}[\ell]=
    {\mathbf{\hat{H}}_{\text{e},k}}[\ell]+{\mathbf{{E}}_{\text{B},k}}[\ell]+
    \sum^{N_{}}_{n=1}{\nu}_n {{\mathbf{E}_{\text{c},k,n}}}[\ell],
\end{equation}
where 
\begin{align}
   \mathbf{\hat{H}}_{\text{e},k}[\ell]&=\mathbf{\hat{H}}_{\text{B},k}[\ell]+\sum^{N_{}}_{n=1}{\nu}_n{{\mathbf{\hat{H}}}_{\text{c},{k,{n}}}}[\ell].
\end{align}
Note that considering the channel estimation errors, the received symbols given by \eqref{1} can be expressed as follows

\begin{align}\label{1e}
   \notag \mathbf{y}_k[\ell]=&\left({\mathbf{\hat{H}}_{\text{e},k}}[\ell]+{\mathbf{{E}}_{\text{B},k}}[\ell]+
    \sum^{N_{}}_{n=1}{\nu}_n {\mathbf{E}_{\text{c},k,{n}}}[\ell]\right)\\
    &\times \sum_{u=1}^{K} {\mathbf{P}}_{u}[\ell]\mathbf{s}_u[\ell]+\boldsymbol{{\eta}}_k[\ell].
\end{align}
At reception, the user $k$ estimates its symbols $\hat{\mathbf{s}}_k[\ell]$ at subcarrier $\ell$ by applying the linear filter $\mathbf{W}_{k}[\ell]\in \mathbb{C}^{N_{{\text{s},k}}[\ell]\times N_{\text{r}}}$, that is, $\hat{\mathbf{s}}_k[\ell]=\mathbf{W}_{k}[\ell]\mathbf{y}_k[\ell]$ (see \Cref{Fig001}). 

The previous imperfect \ac{CSI} model has been developed from a transmitting perspective. However, the same \ac{CSI} model will be considered when designing the receiving filters. Although this is a simplifying assumption, it is a conservative one since we can expect that the \ac{CSI} estimation quality at the receivers will be better than that at the transmitter. 
Considering imperfect \ac{CSI} at the receivers requires the implementation of compensation techniques, similar to those described in \cite{6144751}, to enable coherent detection at the receivers.
We assumed that the estimations of the channels (the direct channels and the cascaded channels) are sent back from the users to the BS through the feedback channel. In practical scenarios, systems with limited rate \ac{CSI} feedback suffers from erroneous \ac{CSI} at the transmitter, which is why we assume imperfect CSI at both sides. The delay effect of the feedback channel is disregarded because is relatively easy to correct (\cite{5482105}) but at the expense of needlessly complicating our notation.

\begin{figure}[h!]
\vspace{-2mm}
\centering
\includegraphics[width=0.98\linewidth]{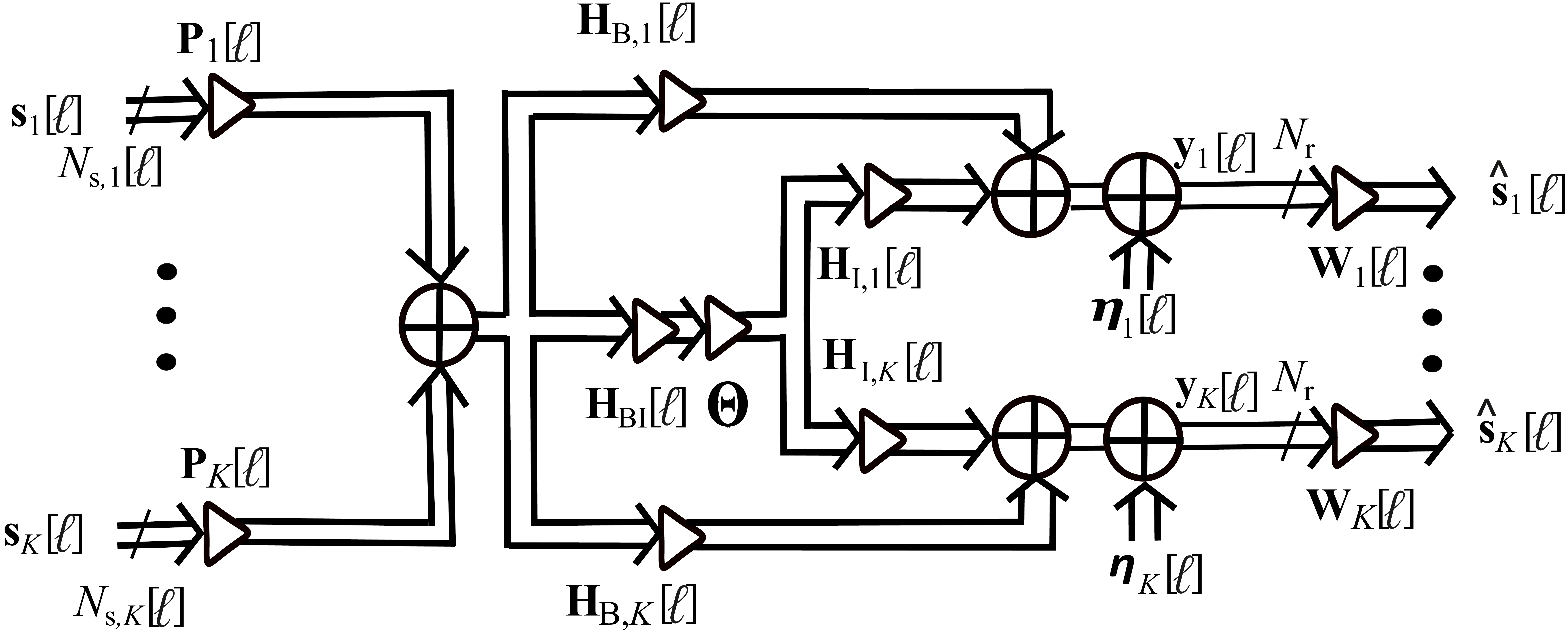}
\vspace{-3.0mm}
\caption{Multistream wideband downlink \ac{MU} \ac{IRS}-aided \ac{mmWave} \ac{MIMO} system at subcarrier $\ell$.}\label{Fig001}
\end{figure}

\Cref{Tsm} summarizes the main system model parameters and variables considered in this multistream wideband downlink \ac{MU} \ac{IRS}-aided \ac{mmWave} \ac{MIMO} system under imperfect \ac{CSI}.
\begin{table}[h!]
\centering
\caption{{Downlink system model parameters.}}\label{Tsm}
	\setlength{\tabcolsep}{5pt}
	\def\arraystretch{1.15}
 \vspace{-2mm}
\begin{tabular}{r|l}
\hline
 {\textbf{Parameter}} & {\textbf{Setting}}  \\ \hline
Number of antennas at the BS&$N_{\text{t}}$\\
Number of users&$K$  \\ 
Number of antennas per users &$N_{\text{r}}$ \\  
Number of subcarriers &$L$\\
Number of streams per user at subcarrier $\ell$ & $N_{\text{s},k}[\ell]$\\
Vector of the $N_{\text{s},k}[\ell]$ symbols at subcarrier $\ell$  &$\mathbf{s}_k[\ell]$\\
 $k$-th \ac{BS} precoder at subcarrier $\ell$ &${\mathbf{P}_{}}_{k}^{}[\ell]$\\
Power available at subcarrier $\ell$ &$P_{\text{T}}{}[\ell]$\\
\ac{BS}-User direct channel  at subcarrier $\ell$&${\mathbf{{H}}^{}_{\text{B},{k}}}[\ell]$\\
\ac{BS}-\ac{IRS} channel subcarrier $\ell$&${\mathbf{{H}}^{}_{\text{BI}}}[\ell]$\\
\ac{IRS}-User channel at subcarrier $\ell$&${\mathbf{{H}}^{}_{\text{I},k}}[\ell]$\\
Frequency-flat \ac{IRS} phase shift matrix & $\mathbf{\Theta}$\\
Main diagonal of \ac{IRS} phase shift matrix & $\boldsymbol{\nu}$\\
Vector of \ac{AWGN} at the $k$-th user at subcarrier $\ell$ & $\boldsymbol{\eta}_k[\ell]$\\
 $k$-th user filter at subcarrier $\ell$ &${\mathbf{W}{}}_{k}^{}[\ell]$\\
 $N_{\text{s},k}[\ell]$ estimated symbols at subcarrier $\ell$& $\hat{\mathbf{s}}_k[\ell]$\\
Total number of streams at subcarrier $\ell$ & $N_{\text{s}}[\ell]$\\
Total number of data streams & $N_{\text{s}}$\\
Covariance matrices of the errors in the downlink & ${\mathbf{C}_{{\eta}}}_k$\\
\hline
\end{tabular}
\end{table}
\subsection{Channel model}
The channels for the different \ac{mmWave} system links are assumed to be time-dispersive. The channel response matrices at the $m$-th delay tap with $m\in\lbrace0,\ldots,L_{\text{D}}-1\rbrace$, where $L_{\text{D}}$ is the maximum number of delay taps, are considered as follows \cite{sayeed_deconstructing_2002,schniter2014channel,9374961}  
\begin{equation}\label{canal_bw}
\B{H}^{}_{\text{temp}}[m]\hspace{-0.0mm}=\hspace{-0.0mm}\gamma \hspace{-0mm} \sum_{j=1}^{{N_{\text{path}}}_{}}
\hspace{-0mm}{\beta_{j}{p_{\text{rc}}}_{}\hspace{-0.0mm}\left(mT_{s}\hspace{-0.0mm}-\hspace{-0.4mm}\tau_{j}\right)\hspace{-0.5mm} \B{a}_{\text{r}}^{}\hspace{-0.0mm}(\phi_{j}^{\text{r}},\psi_{j}^{\text{r}}) 
\B{{a}_{\text{t}}}_{}^{*}\hspace{-0.0mm}(\phi_{j}^{\text{t}},\psi_{j}^{\text{t}}) 
},
\end{equation}
where $N_{\text{path}}$ is the number of channel paths, ${p_{\text{rc}}}_{}(t)$ represents the raised cosine pulse-shaping filter, $\tau_{j}$ is the relative delay for the $j$-th path, $T_{s}$ is the sampling period, $\gamma=\sqrt{N_{\text{t}}N_{\text{r}}/{{N_{\text{path}}}_{}}}$  is a power normalization factor and $\beta_{j}$ stands for the complex path gain for the $j$-th channel path. The term $\phi^{\text{t}}(\psi^{\text{t}})$ stands for the azimuth (elevation) \ac{AoD} and $\phi^{\text{r}}(\psi^{\text{r}})$ are the azimuth (elevation) \ac{AoA}. 
 In the frequency domain, the channel response \eqref{canal_bw} can be represented as \cite{heath_overview_2016}
\begin{equation}
\begin{split}   
\mathbf{H}^{}[\ell]
&=\sum_{m=0}^{L_{\text{D}}-1}\mathbf{H}^{}_{\text{temp}}[m]{e}^{\operatorname{j}2\pi m(\ell-1)/L}\\&= 
\sum_{j=1}^{{N_{\text{path}}}_{}}\beta^{}_{j}[\ell]\B{a}^{}_{\text{r}}[\ell](\phi_{j}^{\text{r}},\psi_{j}^{\text{r}}) 
{\B{a}_{\text{t}}}^{*}_{}[\ell](\phi_{\operatorname{j}}^{\text{t}},\psi_{j}^{\text{t}}), 
\end{split}   
\end{equation}
where $\ell\in\lbrace{1,\ldots,L}\rbrace$ and   $\beta^{}_{j}[\ell]$ represents the $j$-th path gain at subcarrier $\ell$ and corresponds to $$\beta^{}_{j}[\ell]=\gamma\beta_{j}\sum_{m=0}^{L_{\text{D}}-1}{p_{\text{rc}}}_{}(mT_{\text{s}}-\tau_{n}){e}^{j2\pi m(\ell-1)/L}.$$

We assume \acp{UPA} with dimensions $N_\text{a} \times N_\text{b} $ at both communication ends. Therefore, the array response vectors $\B{a}^{}_{\text{t}}(\phi^{\text{t}},\psi^{\text{t}})$ and $\B{a}^{}_{\text{r}}(\phi^{\text{r}},\psi^{\text{r}})$ have the form \cite{balanis2015antenna}
\begin{align*}
\notag \B{a}^{}_{\text{}}\left(\phi,\psi\right)
=\frac{1}{{\sqrt{N_{\text{a}}N_{\text{b}}}}}\Big[&
1,e^{\operatorname{j}\frac{2\pi}{f_{\ell}}c\;d \; (\text{sin}\;\phi\; \text{sin}\;\psi + \text{cos}\;\psi)},
\ldots,\\
& e^{\operatorname{j}\frac{2\pi}{f_{\ell}}c\;d  (\left(  {N_{\text{a}}}-1\right)\;\text{sin}\;\phi\; \text{sin}\;\psi + ({N_{\text{b}}}-1)\text{cos}\;\psi)} \Big]^{\operatorname{T}}
\end{align*}
 where $f_{\ell}$ is the frequency corresponding to the $\ell$-th subcarrier, $c$ is the speed of light, and $d$ is the inter antenna spacing which is often set to $\lambda/2$, with $\lambda=\frac{c}{f_{\text{c}}}$ the wavelength corresponding to the carrier frequency $f_{\text{c}}$. 

\subsection{Achievable sum-rate}
 
 Considering the described scenario (see also \Cref{Fig001}) and imperfect \ac{CSI} model, we aim at determining the precoders $\mathbf{P}^{}_k[\ell], \forall k,\ell$, the receiving linear filters $\mathbf{W}^{}_{k}[\ell],\forall k,\ell$, and the \ac{IRS} phase shift matrix $\mathbf{\Theta}^{}$ which maximize the achievable sum-rate given by\footnote{{While robust design and statistical approaches under imperfect CSI in the transmitter often involve considering channel outages, our specific focus is on rate maximization, as exemplified in works such as \cite{9110587} and \cite{gonzalez2018qos}.}}
\begin{align}\label{RW}
R_{\text{sum}}=& \notag \sum_{\ell=1}^{L}\sum_{k=1}^{K}\text{log}_{2}\;\text{det}\;\Big(\mathbf{I}_{{N_{\text{s},k}}[\ell]}+\\
& \mathbf{X}_k^{-1}[\ell]\mathbf{W}_k[\ell]
{\mathbf{H}_{\text{e},k}}[\ell]\mathbf{P}_{k}[\ell]{\mathbf{P}}^{*}_{k}[\ell] {\mathbf{H}^*_{\text{e},k}}[\ell]\mathbf{W}^*_k[\ell]\Big)
\end{align} 
with \begin{align}
\notag
\mathbf{X}_k[\ell]\hspace{-1mm}=&\sum_{i\neq k} 
\mathbf{W}_k[\ell]{\mathbf{H}_{\text{e},k}}[\ell]{\mathbf{P}^{}}_{i}[\ell]
{\mathbf{P}}_{i}^{*}[\ell]{\mathbf{H}^*_{\text{e},k}}[\ell]\mathbf{W}^*_k[\ell]\\&
+\mathbf{W}_k[\ell]\mathbf{C}_{\eta_k}\mathbf{W}^*_k[\ell],
\end{align}
while fulfilling the overall power constraint ($P_{\text{T}}$). Recall that ${\mathbf{H}_{\text{e},k}}[\ell]$ represents the equivalent \ac{BS}-\ac{IRS}-User channel for user $k$ at subcarrier $\ell$. Accordingly, we formulate the following optimization problem\footnote{{Note that a similar optimization problem could be formulated by considering an individual minimum data rate per user. This would entail changes in the kind of MAC-BC and BC-MAC dualities (second or third kind dualities \cite{hunger2008mse}) explained below.}}}

\begin{align}\label{op1}
&\underset{\mathbf{\Theta}^{},\mathbf{P}_k[\ell],\mathbf{W}_{k}[\ell],\forall k,\ell}{\text{arg}\;\text{max}}\;\;R_{\text{sum}}
\\&
\text{s.t.}\;{\sum_{k=1}^{K} \text{tr}({\mathbf{P}_{k}[\ell]}{\mathbf{P}^{*}_{k}[\ell]}})\leq P_{\text{T}}[\ell], \forall \ell,\; \mathbf{\Theta}^{}\in\mathcal{D} \notag
.\;\notag
\end{align}
The solution to \eqref{op1} is computationally intractable mainly because of the coupling of the variables $\mathbf{P}_k$ and $\mathbf{\Theta}$, the non-convex nature of the cost function and the non-convex constraints on the \ac{IRS} matrix, i.e., unit magnitude for each diagonal entry. 

Furthermore, the \ac{IRS} phase shift matrix is common to all the users and subcarriers which makes the solution of \eqref{op1} even more difficult since the common \ac{IRS} must be designed to simultaneously modify all the paths of the user channels at the $L$ subcarriers.
Note that the exhaustive search technique is required to find an optimal solution, but its impracticality arises due to the high computational complexity, even for moderate-sized systems. As a result, achieving a globally optimal solution for these complex wideband \ac{MU} MIMO IRS-aided systems is not feasible.

\section{{MMSE} Approach}\label{Section III}
We reformulate the optimization problem \eqref{op1} as an \ac{MSE} minimization problem since minimizing the sum-\ac{MSE}  maximizes a lower bound of the system capacity \cite{christensen2008weighted}. 
Weighted MMSE could be used instead of sum MSE minimization, but this would require an additional inner iterative loop in our design algorithm to optimize the MSE weights \cite{christensen2008weighted}. Some performance improvement is to be expected but at the expense of a significant increase in computational complexity and a slowdown in convergence.

Let us elaborate the system model under imperfect \ac{CSI} for the $k$-th user symbols at the $\ell$-th subcarrier after the linear filtering at reception, i.e.,
\begin{align}\label{recx}
  \notag\mathbf{\hat{s}}_k[\ell]=&\mathbf{W}_k[\ell]\left({\mathbf{\hat{H}}_{\text{e},k}}[\ell]+{\mathbf{{E}}_{\text{B},k}}[\ell]+
    \sum^{N_{}}_{n=1}{\nu}_n {{\mathbf{E}_{\text{c},k,n}}}[\ell]\right)\\
    &\times\sum_{u=1}^{K} {\mathbf{P}}_{u}[\ell]\mathbf{s}_u[\ell]+\mathbf{W}_k[\ell]\boldsymbol{{\eta}}_k[\ell].
\end{align}
Thus, the downlink \ac{MSE} for the $k$-th user at subcarrier $\ell$ is
\begin{align}\label{mmseicsi}
\notag &\text{MSE}^{\text{DL}}_k[\ell]=\mathbb{E}\left[ \Vert\mathbf{s}_k[\ell]-\hat{\mathbf{s}}_k[\ell]\Vert_2^{2}\right] 
\notag \\ &\notag= \text{tr} \Bigg(\mathbf{W}_k[\ell]\Bigg\{ {\mathbf{\hat{H}}_{\text{e},k}}[\ell] \left[ \sum_{i=1}^{K} \mathbf{P}_i[\ell] \mathbf{P}^*_i[\ell]\right]\\
&\notag \times{\mathbf{\hat{H}}^*_{\text{e},k}}[\ell]  +\mathbf{C}_{{\eta}_k}\Bigg\}\mathbf{W}^*_k[\ell]\\
\notag&+ \mathbb{E}\Bigg[\mathbf{W}_k[\ell] \Bigg\{{\mathbf{{E}}_{\text{B},k}}[\ell] \left[ \sum_{i=1}^{K} \mathbf{P}_i[\ell] \mathbf{P}^*_i[\ell]\right]{\mathbf{{E}}^*_{\text{B},k}}[\ell]\Bigg\}\mathbf{W}^*_k[\ell]\Bigg]\\
\notag&+ \mathbb{E}\Bigg[\mathbf{W}_k[\ell] \Bigg\{\sum_{n=1}^N{\mathbf{{E}}_{\text{c},k,n}}[\ell]{\nu}_n \left[ \sum_{i=1}^{K} \mathbf{P}_i[\ell] \mathbf{P}^*_i[\ell]\right]\\
&\notag\times\sum_{n^\prime=1}^N{\mathbf{{E}}^*_{\text{c},k,{n^\prime}}}[\ell]{\nu}_{n^\prime}^*\Bigg\}\mathbf{W}^*_k[\ell]\Bigg]\\
&-\mathbf{W}_k[\ell]{\mathbf{\hat{H}}_{\text{e},k}}[\ell] \mathbf{P}_k[\ell]
-
\mathbf{P}^*_k[\ell]{\mathbf{\hat{H}}^*_{\text{e},k}}[\ell] \mathbf{W}^*_k[\ell]+\mathbf{I}_{{N}_{\text{s},k}}[\ell]\Bigg).
\end{align}
This expression for the downlink \ac{MSE} in terms of the precoders $\mathbf{P}_k[\ell]$, the filters $\mathbf{W}_k[\ell]$, the estimated channels ${\mathbf{\hat{H}}^*_{\text{e},k}}[\ell]$ and the errors ${\mathbf{{E}}_{\text{B},k}}[\ell]$, ${\mathbf{{E}}_{\text{c},k,n}}[\ell],\forall n,k$  can be rewritten as follows
\begin{align}\label{MSEDLd}
&\notag\text{MSE}^{\text{DL}}_k[\ell]= \text{tr}\Bigg(\mathbf{W}_k[\ell]\Bigg\{ {\mathbf{\hat{H}}_{\text{e},k}}[\ell] \left[ \sum_{i=1}^{K} \mathbf{P}_i[\ell] \mathbf{P}^*_i[\ell]\right]\\
&\times\notag{\mathbf{\hat{H}}^*_{\text{e},k}}[\ell] +\mathbf{C}_{{\eta}_k}\Bigg\}\mathbf{W}^*_k[\ell]\Bigg)\\
\notag&+\text{tr}\Bigg(   \sum_{i=1}^{K} \mathbf{P}_i[\ell] \mathbf{P}^*_i[\ell]\Bigg)\text{tr}\Bigg(\mathbf{W}_k[\ell]{\mathbf{C}_{\eta}}_k
\mathbf{W}^*_k[\ell]\Bigg)\frac{L}{P_{\text{T}}}\\
\notag&+\boldsymbol{\nu}^*\text{tr}\Bigg( \sum_{i=1}^{K} \mathbf{P}_i[\ell] \mathbf{P}^*_i[\ell]\Bigg)\text{tr}\Bigg(\mathbf{W}_k[\ell]{\mathbf{C}_{\eta}}_k
\mathbf{W}^*_k[\ell]\Bigg)\boldsymbol{\nu}\frac{L}{P_{\text{T}}}\\
&-\text{tr}\Bigg(\mathbf{W}_k[\ell]{\mathbf{\hat{H}}_{\text{e},k}}[\ell] \mathbf{P}_k[\ell]
-
\mathbf{P}^*_k[\ell]{\mathbf{\hat{H}}^*_{\text{e},k}}[\ell] \mathbf{W}^*_k[\ell]+\mathbf{I}_{{N}_{\text{s},k}[\ell]}\Bigg),
\end{align} 
where the independence of the columns of $\mathbf{E}_{\text{B},k}[\ell]$ and $\mathbf{E}_{\text{c},n,k}[\ell]$ has been exploited. Note that \eqref{MSEDLd} contains the covariance matrices $\frac{L}{P_{\text{T}}}\mathbf{C}_{\eta_k}$ of the error matrices $\mathbf{E}_{\text{B},k}[\ell]$ and $\mathbf{E}_{\text{c},n,k}[\ell]$. Hence, the statistical information of the channel estimation errors is incorporated in \eqref{MSEDLd}. 
Although we assumed the LS method for channel estimation in \Cref{ches}, our proposed solution is not limited to this error model in passive IRS-aided systems. Equation \eqref{MSEDLd} is independent of the specific error models used for aggregated CSI estimation in \eqref{llv} and can be adapted to various channel estimation techniques with different covariance matrices for the errors.

For given precoders $\mathbf{P}_k[\ell]$ and \ac{IRS} phase shift matrix $\mathbf{\Theta}$, the $k$-th user's \ac{MMSE} receiving filter is readily determined as 
\begin{align}\label{mmsefil}
&\hspace{-0.5mm}{\mathbf{W}_{\hspace{-0.5mm}\text{MMSE},k}}\hspace{-0.5mm}[\ell]\hspace{-1mm}=\hspace{-1mm}\mathbf{P}^*_{k}[\ell]{\mathbf{\hat{H}}^*_{\text{e},{k}}}[\ell]
\hspace{-0.5mm}\Big(\hspace{-0.8mm}{\mathbf{\hat{H}}_{\text{e},{k}}}[\ell]
\hspace{-0.5mm}\mathbf{P}_{k}[\ell]\hspace{-0.5mm}\mathbf{P}^*_{k}[\ell]{\mathbf{\hat{H}}^*_{\text{e},{k}}}[\ell]
\hspace{-1mm}+\hspace{-1mm}\mathbf{C}_{\text{IN},k}[\ell]\hspace{-0.9mm}\Big)^{\hspace{-1.5mm}-1}
\end{align}
where  $\mathbf{C}_{\text{IN},k}[\ell]\in \mathbb{C}^{N_\text{r} \times N_\text{r}}$ is the $k$-th interference-plus-noise covariance matrix given by 
\begin{align}\label{cink}
\notag\mathbf{C}_{\text{IN},k}[\ell]=& \sum_{i\neq k} 
{\mathbf{\hat{H}}_{\text{e},k}}[\ell]{\mathbf{P}^{}}_{i}[\ell] {\mathbf{P}}_{i}^{*}[\ell]{\mathbf{\hat{H}}^*_{\text{e},k}}[\ell]+\mathbf{C}_{{\eta}_k} \\
&+(N+1)\sum_{i=1}^K \text{tr}(  \mathbf{P}_i[\ell] \mathbf{P}^*_i[\ell])\mathbf{C}_{{\eta}_k}\frac{L}{P_{\text{T}}}.
\end{align}
Substituting \eqref{mmsefil} into \eqref{recx} leads to the following expression for the downlink \ac{MSE}
\begin{equation}\label{msek}
\text{MSE}^{\text{DL}}_k[\ell]= \text{tr}(\mathbf{I}_{{N}_{\text{s},k}}[\ell]+\mathbf{P}^*_{k}[\ell]{\mathbf{\hat{H}}^*_{\text{e},{k}}}[\ell]\mathbf{C}^{-1}_{\text{IN},k}[\ell]{\mathbf{\hat{H}}_{\text{e},{k}}}[\ell]\mathbf{P}_{k}[\ell])^{-1}.
\end{equation} 

Finally, in order to  determine the wideband precoders $\mathbf{P}_k[\ell]$ and the \ac{IRS} phase-shift matrix $\mathbf{\Theta}$ assuming imperfect \ac{CSI}, we formulate the following downlink \ac{MSE} minimization problem 
\begin{align}\label{opmsek}
&\underset{\mathbf{\Theta}^{},\mathbf{P}_{k}[\ell],\forall k,\ell}{\text{arg}\;\text{min}}\;\;{\sum_{\ell=1}^L\sum_{k=1}^{K}\text{MSE}^{\text{DL}}_{k}[\ell]}
\\&
\text{s.t.}\;\sum_{k=1}^{K} \text{tr}({\mathbf{P}_{k}}[\ell]{\mathbf{P}^{*}_{k}}[\ell])\leq P_{\text{T}}[\ell],\forall \ell,\; \mathbf{\Theta}^{}\in\mathcal{D} \notag
,\;\notag
\end{align}
with $\text{MSE}^{\text{DL}}_k[\ell]$ given by \eqref{msek} and considering $\mathbf{C}_{\text{IN},k}[\ell]$ as in \eqref{cink}. In the ensuing section, this minimization problem will be formulated in the dual \ac{MAC} in order to obtain a more appropriate mathematical structure for the \ac{MSE}. This will lead to a significant reduction in computational complexity.

\section{Proposed Solution}\label{Section IV}
\begin{figure}[h!]
\centering
\includegraphics[width=0.97\linewidth]{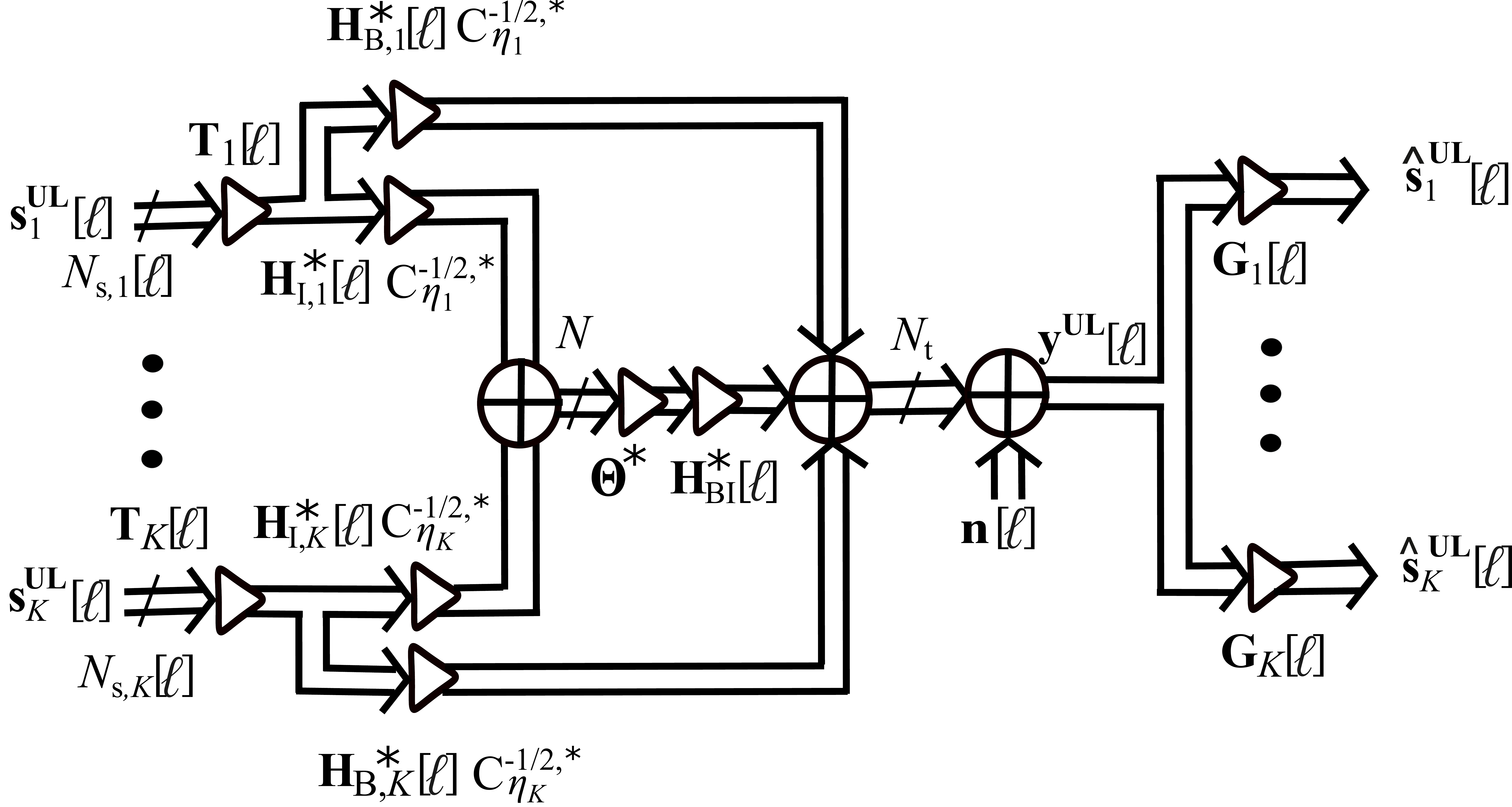}
\vspace{-3.5mm}
\caption{Multistream wideband dual uplink \ac{MU} \ac{IRS}-aided \ac{mmWave} \ac{MIMO} system model at subcarrier $\ell$.}\label{Fig002}
\end{figure}

In this section, we will exploit the \ac{BC}-\ac{MAC} duality \cite{hunger2008mse,endeshaw2009mse} to solve \eqref{opmsek}. We start by defining the virtual dual \ac{MAC} model for the considered scenario. Let $\mathbf{s}^{\text{UL}}_k[\ell]$ with $\mathbb{E}[\mathbf{s}^{\text{UL}}_k[\ell]\mathbf{s}^{\text{UL,*}}_k[\ell]]=\mathbf{I}_{N_{\text{s},k}[\ell]}$ be the zero-mean vector of uplink symbols transmitted by the users over subcarrier $\ell$ in the dual \ac{MAC}. The corresponding uplink symbols received at the \ac{BS} will be (see \Cref{Fig002})
\begin{equation}\label{yul}
\mathbf{y}^{\text{UL}}[\ell]=  \sum^{K}_{k=1}  {\mathbf{H}^{*}_{\text{e},k}}[\ell] \mathbf{C}_{\eta_k}^{-1/2,*}\mathbf{T}_k[\ell]\mathbf{s}^{\text{UL}}_k[\ell]+\mathbf{n}[\ell],
\end{equation}
where $\mathbf{T}_k[\ell]\in\mathbb{C}^{N_{\text{r}}\times N_{\text{s},k}[\ell]}$ and ${\mathbf{H}^*_{\text{e}_k}}[\ell]\mathbf{C}_{\eta_k}^{-1/2,*}$ are the $k$-th user precoder and the equivalent channel response in the dual \ac{MAC}, respectively, for subcarrier $\ell$. The noise at subcarrier $\ell$ introduced at reception is represented by $\B{n}[\ell]\sim \mathcal{N}_{\mathbb{C}} (0, \B{I}_{N_{\text{t}}})$. Note that
\begin{align}
{\mathbf{H}^{*}_{\text{e},k}}[\ell]=\mathbf{H}^*_{\text{BI}}[\ell]\mathbf{\Theta}^*{\mathbf{H}^*_{\text{I},k}}[\ell]+{\mathbf{H}^*_{\text{B},k}}[\ell]
\end{align}
is the conjugate transpose of the joint cascaded and direct channel matrices for the $k$-th user at subcarrier $\ell$ given by \eqref{mm}.
We now introduce  $\mathbf{G}_{k}[\ell]\in\mathbb{C}^{N_{\text{s},k}[\ell]\times N_{\text{t}}},\;\forall k$ which are the linear filters used at reception in the dual \ac{MAC} to estimate the incoming uplink symbols for user $k$ and subcarrier $\ell$, i.e., 
\begin{align}\label{kup}
	\notag\hat{\mathbf{s}}^{\text{UL}}_k[\ell]&= \mathbf{G}_k[\ell] \mathbf{y}^{\text{UL}}[\ell] \\
	&= \mathbf{G}_k[\ell]\sum^K_{i=1}{\mathbf{H}^{*}_{\text{e},i}}[\ell]\mathbf{C}_{\eta_k}^{-1/2,*}\mathbf{T}_i[\ell]\mathbf{s}^{\text{UL}}_i[\ell]+\mathbf{G}_k[\ell]\mathbf{n}[\ell].
\end{align}
By considering imperfect \ac{CSI} and a similar approach to that in \Cref{ches}, the uplink symbol expression \eqref{kup} can be rewritten as [cf. \eqref{heqv}]
\begin{align}\label{kupicsi}
	\notag \hat{\mathbf{s}}^{\text{UL}}_k[\ell]=& \mathbf{G}_k[\ell]\sum^K_{i=1}
		\Bigg({\mathbf{\hat{H}}^{*}_{\text{e},i}}[\ell]+
	{\mathbf{E}^{*}_{\text{B},i}}[\ell]+
	\sum^{N_{}}_{n=1}{{\mathbf{E}^*_{\text{c},k,n}}}[\ell]{\nu}^*_n
	\Bigg)\\
	&
	\times \mathbf{C}_{\eta_k}^{-1/2,*} \mathbf{T}_i[\ell]\mathbf{s}^{\text{UL}}_i[\ell]+\mathbf{G}_k[\ell]\mathbf{n}[\ell].
\end{align}

We next determine the uplink \ac{MSE} between the sent and the estimated symbols, i.e., $\text{MSE}^{\text{UL}}_k[\ell] =\mathbb{E}\left[ \Vert\mathbf{s}^{\text{UL}}_k[\ell]-\hat{\mathbf{s}}^{\text{UL}}_k[\ell]\Vert_2^{2}\right]$. 
This \ac{MSE} per user in the dual uplink at subcarrier $\ell$, when considering imperfect \ac{CSI}, is given by
\begin{align}
\notag\text{MSE}^{\text{UL}}_k[\ell]&= \text{tr}\Bigg(\mathbf{G}_k[\ell]\Bigg\{  \Bigg[ \sum_{i=1}^{K} {\mathbf{\hat{H}}^*_{\text{e},i}}[\ell] \mathbf{C}_{\eta_k}^{-1/2,*} \mathbf{T}_i[\ell] \\
&\notag\times\mathbf{T}^*_i[\ell]\mathbf{C}_{\eta_k}^{-1/2}{\mathbf{\hat{H}}_{\text{e},i}}[\ell]\Bigg]+\mathbf{I}_{N_{\text{t}}}\Bigg\}\mathbf{G}^*_k[\ell]\Bigg)\\
\notag&+\sum_{i=1}^{K}\text{tr}\Bigg(\frac{L}{P_{\text{T}}} \mathbf{T}_i[\ell] \mathbf{T}^*_i[\ell]\Bigg)\text{tr}\Bigg(\mathbf{G}_k[\ell]
\mathbf{G}^*_k[\ell]\Bigg)(N+1)\\\notag
&-\text{tr}\Bigg(\mathbf{G}_k[\ell]{\mathbf{\hat{H}}^*_{\text{e},k}}[\ell]\mathbf{C}_{\eta_k}^{-1/2,*} \mathbf{T}_k[\ell]
\\&-
\hspace{2mm}\mathbf{T}^*_k[\ell]\mathbf{C}_{\eta_k}^{-1/2}{\mathbf{\hat{H}}_{\text{e},k}}[\ell]\mathbf{G}^*_k[\ell]+\mathbf{I}_{{N}_{\text{s},k}[\ell]}\Bigg),
\end{align} 
since $\boldsymbol{\nu}^*\boldsymbol{\nu}=N$. The \ac{MMSE} filter in the uplink is given by
\begin{align}\label{mmsefuimper}
\notag& {\mathbf{G}_{\text{MMSE},k}}[\ell]=\mathbf{T}^*_k[\ell]\mathbf{C}_{\eta_k}^{-1/2}{\mathbf{\hat{H}}_{\text{e}_k}}[\ell]\\\notag
&\times \Big({\mathbf{\hat{H}}^*_\text{e}} [\ell]\mathbf{C}_{\eta_k}^{-1/2,*}\mathbf{T}[\ell]\mathbf{T}^*[\ell]\mathbf{C}_{\eta_k}^{-1/2}{\mathbf{\hat{H}}_\text{e}}[\ell] +\mathbf{I}_{N_{\text{t}}}
\\ 
& + (1+N)\;\sum_{i=1}^{K}\text{tr}\Big(\frac{L}{P_{\text{T}}} \mathbf{T}_i[\ell] \mathbf{T}^*_i[\ell]\Big)\mathbf{I}_{N_{\text{t}}}
\Big)^{-1},
\end{align}
where $\mathbf{T}[\ell]=\text{blkdiag}\left(\mathbf{T}_1[\ell],\ldots,\mathbf{T}_K[\ell]\right)$ is a block diagonal matrix that comprises all the dual uplink user precoders, $\mathbf{C}_{}=\text{blkdiag}(\mathbf{C}_{\eta_1},\ldots,\mathbf{C}_{\eta_K})\in \mathbb{C}^{{N_{\text{r}}} K\times {N_{\text{r}}} K }$ is another block diagonal matrix stacking the inverse noise covariance matrices corresponding to all the users, and $\mathbf{\hat{H}}^{*}_{\text{e}}[\ell]=\left[{\mathbf{\hat{H}}^{*}_{\text{e},1}}[\ell],\ldots, {\mathbf{\hat{H}}^{*}_{\text{e},K}}[\ell]\right]$ stacks all the equivalent channel response matrices. 
 The \ac{MSE} per user achieved at subcarrier $\ell$ when employing this \ac{MMSE} receiving filter is

\begin{align}\label{MSEup}
\notag\text{MSE}^{\text{UL}}_k[\ell]&= \text{tr}\Bigg[\Bigg(\mathbf{I}_{{N}_{\text{s},k}[\ell]}+ \mathbf{T}^*_k[\ell]\mathbf{C}_{\eta_k}^{-1/2}{\mathbf{\hat{H}}_{\text{e},k}}[\ell]\\
\times & \notag \Bigg(\sum_{i\neq k}^{K} {\mathbf{\hat{H}}^*_{\text{e},i}}[\ell]\mathbf{C}_{\eta_i}^{-1/2,*} \mathbf{T}_i[\ell] \mathbf{T}^*_i[\ell]\mathbf{C}_{\eta_i}^{-1/2}{\mathbf{\hat{H}}_{\text{e},i}}[\ell] \\+&\notag\mathbf{I}_{N_{\text{t}}}+(1+N)\;\sum_{i=1}^{K}\text{tr}\Big(\frac{L}{P_{\text{T}}} \mathbf{T}_i[\ell] \mathbf{T}^*_i[\ell]\Big)\mathbf{I}_{N_{\text{t}}}\Bigg)^{-1}\\
\times&\mathbf{\hat{H}}^*_{\text{e},k}[\ell] \mathbf{C}_{\eta_k}^{-1/2,*}\mathbf{T}_k[\ell]\Bigg)^{-1}\Bigg],
\end{align}
Note that this expression only depends on the uplink precoders in the dual \ac{MAC} $\mathbf{T}_k[\ell]$ and the \ac{IRS} phase-shift matrix $\mathbf{\Theta}$.

We next rewrite the received uplink symbols at subcarrier $\ell$ given by \eqref{yul} by means of the following more compact expression (cf. the definitions below \eqref{mmsefuimper})
\begin{equation}\label{equl}
\mathbf{y}^{\text{UL}}[\ell]=\mathbf{H}^{*}_{\text{e}}[\ell]\mathbf{C}_{\eta}^{-1/2,*}\mathbf{T}[\ell]\mathbf{s}^{\text{UL}}[\ell]+\mathbf{n}[\ell],
\end{equation}
where $\mathbf{s}^{\text{UL}}[\ell]=[\mathbf{s}^{\text{UL},{\operatorname{T}}}_1[\ell],\ldots,\mathbf{s}^{\text{UL},{\operatorname{T}}}_K[\ell]]^{\operatorname{T}}$ is a vector that gathers all the uplink user symbols sent over the $\ell$-th subcarrier. According to \eqref{mm}, this latter matrix is related to the different channel responses in the signal model as follows
\begin{equation}\label{cmm}
\mathbf{H}^{*}_{\text{e}}[\ell]=(\mathbf{H}^*_{\text{BI}}[\ell]\mathbf{\Theta}^*\mathbf{H}^*_{\text{I}}[\ell]+{\mathbf{H}^*_{\text{B}}}[\ell])\in\mathbb{C}^{N_{\text{t}}\times N_{\text{r}}K},
\end{equation} 
where $\mathbf{H}^{*}_{\text{I}}[\ell]=\left[{\mathbf{H}^{*}_{\text{I},1}}[\ell],\ldots, {\mathbf{H}^{*}_{\text{I},K}}[\ell]\right]\in \mathbf{C}_{\eta}^{-1/2,*}$, and $\mathbf{H}^{*}_{\text{B}}[\ell]=\left[{\mathbf{H}^{*}_{\text{B},1}}[\ell],\ldots, {\mathbf{H}^{*}_{\text{B},K}}[\ell]\right]\in \mathbb{C}^{N_{\text{t}}\times N_{\text{r}}K}$.
 
Substituting the frequency selective equivalent channel responses given by \eqref{cmm} into  \eqref{equl} leads to 
 \begin{equation}
\mathbf{y}^{\text{UL}}[\ell]=(\mathbf{H}^*_{\text{BI}}[\ell]\mathbf{\Theta}^*\mathbf{H}^*_{\text{I}}[\ell]+\mathbf{H}^{*}_{\text{B}}[\ell])\mathbf{C}_{\eta}^{-1/2,*}\mathbf{T}[\ell]\mathbf{s}^{\text{UL}}[\ell]+\mathbf{n}[\ell].
 \end{equation}
Note that $\mathbf{T}[\ell]\in\mathbb{C}^{N_{\text{r}}K\times N_{\text{s}}[\ell]}$ and $\mathbf{s}^{\text{UL}}[\ell]\in\mathbb{C}^{N_{\text{s}}[\ell]}$ with $N_{\text{s}}[\ell]=\sum_{k=1}^K N_{\text{s},k}[\ell]$, i.e., the dimensions of $\mathbf{T}[\ell]$ and $\mathbf{s}^{\text{UL}}[\ell]$ depend on the number of served data streams per user. 

The estimated uplink data symbols corresponding to the $K$ users at subcarrier $\ell$ according to the dual \ac{MAC} signal model can hence be defined as
\begin{align}\label{allup}
\notag\hat{\mathbf{s}}^{\text{UL}}[\ell]=&\mathbf{G}[\ell](\mathbf{H}^*_{\text{BI}}[\ell]\mathbf{\Theta}^*\mathbf{H}^*_{\text{I}}[\ell]+\mathbf{H}^{*}_{\text{B}}[\ell])\\&\times\mathbf{C}_{\eta}^{-1/2,*}\mathbf{T}[\ell]\mathbf{s}^{\text{UL}}[\ell]+\mathbf{G}[\ell]\mathbf{n}[\ell],
\end{align}
where $\mathbf{G}[\ell]=\left[\mathbf{G}^*_1[\ell],\ldots,\mathbf{G}^*_K[\ell]\right]^*$ is the overall receive filter that comprises the filters $\mathbf{G}_k[\ell],\; \forall k$ which estimate all the uplink symbols at subcarrier $\ell$. Note that $\hat{\mathbf{s}}^{\text{UL}}[\ell]$ collects all the estimated user symbols $\hat{\mathbf{s}}_k^{\text{UL}}[\ell],\forall k$, which can also be obtained through \eqref{kup}. Considering that the \ac{CSI} is imperfect, the estimated uplink user symbols $\hat{\mathbf{s}}^{\text{UL}}[\ell]$ can be represented as follows
\begin{align}\label{upicsi}
	&\notag \hat{\mathbf{s}}^{\text{UL}}[\ell]= \mathbf{G}[\ell]
		\Bigg({\mathbf{\hat{H}}^{*}_{\text{B}}}[\ell]+
	{\mathbf{E}^{*}_{\text{B}}}[\ell]+
	\sum^{N_{}}_{n=1}\Big({{\mathbf{\hat{H}}}^*_{\text{c},{{n}}}}[\ell]+{{\mathbf{E}^*_{\text{c},{n}}}}[\ell]\Big){\nu}^*_n
	\Bigg)\\
	&
	\times \mathbf{C}_{\eta}^{-1/2,*} \mathbf{T}[\ell]\mathbf{s}^{\text{UL}}[\ell]+\mathbf{G}[\ell]\mathbf{n}[\ell],
\end{align}
with $\mathbf{E}^{*}_{\text{B}}[\ell]=\left[{\mathbf{E}^{*}_{\text{B},1}}[\ell],\ldots, {\mathbf{E}^{*}_{\text{B},K}}[\ell]\right]\in \mathbb{C}^{N_{\text{t}}\times N_{\text{r}}K}$ and  ${\mathbf{E}^{*}_{\text{c},n}}[\ell]=\left[{\mathbf{{E}}_{{\text{c},{1},n}}}[\ell],\ldots, {\mathbf{{E}}_{{\text{c},K,n}}}[\ell]\right]\in \mathbb{C}^{N_{\text{t}}\times N_{\text{r}}K}$, $
\notag {{\mathbf{\hat{H}}}^*_{\text{c},{{n}}}}[\ell]=\Bigg[ {{\mathbf{\hat{H}}}^*_{\text{c},1,{n}}}[\ell],\ldots, {{\mathbf{\hat{H}}}^*_{\text{c},K,{n}}}[\ell]\Bigg] \in \mathbb{C}^{N_{\text{t}}\times N_{\text{r}}K},\forall n.
$
The \ac{MMSE} receive filter using this compact formulation is given by
\begin{align}\label{mmsefuk}
\notag& {\mathbf{G}_{\text{MMSE}}}[\ell]=\mathbf{T}^*[\ell]
\mathbf{C}_{\eta}^{-1/2}
{\mathbf{\hat{H}}_{\text{e}_k}}[\ell]\\\notag
&\times \Big({\mathbf{\hat{H}}^*_\text{e}} [\ell]\mathbf{C}_{\eta}^{-1/2,*}\mathbf{T}[\ell]\mathbf{T}^*[\ell]\mathbf{C}_{\eta}^{-1/2}{\mathbf{\hat{H}}_\text{e}}[\ell] +\mathbf{I}_{N_{\text{t}}}
\\ 
& + (1+N)\;\sum_{i=1}^{K}\text{tr}\Big(\frac{L}{P_{\text{T}}} \mathbf{T}_i[\ell] \mathbf{T}^*_i[\ell]\Big)\mathbf{I}_{N_{\text{t}}}
\Big)^{-1},
\end{align}
and the overall uplink \ac{MSE} achieved at subcarrier $\ell$ when employing this \ac{MMSE} receiving filter is

\begin{align}\label{MSEupc}
\notag&\text{MSE}^{\text{UL}}[\ell]= \text{tr}\Bigg[\Bigg(\mathbf{I}_{{N}_{\text{s}}[\ell]}+ \mathbf{T}^*[\ell]\mathbf{C}_{\eta}^{-1/2}{\mathbf{\hat{H}}_{\text{e}}}[\ell]\\
\times&\notag\Bigg( %
\mathbf{I}_{N_{\text{t}}}+(1+N)\;\sum_{i=1}^{K}\text{tr}\Big(\frac{L}{P_{\text{T}}} \mathbf{T}_i[\ell] \mathbf{T}^*_i[\ell]\Big)\mathbf{I}_{N_{\text{t}}}\Bigg)^{-1}\\
\times&\mathbf{\hat{H}}^*_{\text{e}}[\ell] \mathbf{C}_{\eta}^{-1/2,*}\mathbf{T}[\ell]\Bigg)^{-1}\Bigg].
\end{align}

According to \cite{endeshaw2009mse,5670866}, the filters and precoders in the downlink are simply related to their counterparts in the dual \ac{MAC} as follows:
\begin{equation}\label{dualp}
\mathbf{P}_{}[\ell]=\xi[\ell]\mathbf{G}^*_{}[\ell], \end{equation}
\begin{equation}\label{dualw}
\mathbf{W}[\ell]=\xi^{-1}_{}[\ell]\mathbf{T}^*[\ell] \mathbf{C}_{\eta}^{-1/2},
\end{equation} 
with $\xi[\ell] \in \mathbb{R}$, given by
\begin{align}\label{xi}
\xi[\ell]=\sqrt{\frac{P_{\text{T}}[\ell]}{\sum_{k=1}^K \|\mathbf{G}_k[\ell]\|_F^2}},
\end{align}
where $P_{\text{T}}[\ell]$ is the power allocated at subcarrier $\ell$. On the other hand, the \ac{MAC}-\ac{BC} dual relationship can be easily established as
\begin{equation}\label{dualpu}
\mathbf{G}_{}[\ell]={\zeta}^{-1}[\ell]\mathbf{P}^*_{}[\ell], \end{equation}
\begin{equation}\label{dualwu}
\mathbf{T}[\ell]=\mathbf{C}_{\eta}^{-1/2,*}\mathbf{W}^*[\ell]\zeta_{}[\ell],
\end{equation} with $\zeta[\ell] \in \mathbb{R}$ given by
\begin{align}\label{xiu}
\zeta[\ell]=\sqrt{\frac{P_{\text{T}}[\ell]}{\sum_{k=1}^K \|\mathbf{W}_k[\ell]\|_F^2}}.
\end{align}
Because of duality, the same sum-\ac{MSE} is achieved in the downlink when using \eqref{dualp} and \eqref{dualw} to obtain the wideband downlink precoders and filters from the wideband uplink filters and precoders. \Cref{UTsm} summarizes the main system model parameters and variables for the dual \ac{MAC} system. 
 \begin{table}[h!]
\centering
\caption{{Dual \ac{MAC} system model parameters.}}\label{UTsm}
	\setlength{\tabcolsep}{5pt}
	\def\arraystretch{1.15}
   \vspace{-2mm}
\begin{tabular}{r|l}
\hline
 {\textbf{Parameter}} & {\textbf{Setting}}  \\ \hline

Precoder of $k$-th user at subcarrier $\ell$ &${\mathbf{T}_{}}_{k}^{}[\ell]$\\
 (\ac{BC}-\ac{MAC}) frequency flat IRS phase shift matrix & $\mathbf{\Theta}^*$\\
User-\ac{BS} direct channel of $k$-th user at subcarrier $\ell$&${\mathbf{{H}}^{*}_{\text{B},{k}}}[\ell]\mathbf{C}_{\eta}^{-1/2,*}$\\
\ac{BS}-\ac{IRS} channel at subcarrier $\ell$&${\mathbf{{H}}^{*}_{\text{BI}}}[\ell]$\\
\ac{IRS}-User channel of $k$-th user at subcarrier $\ell$&${\mathbf{{H}}^{*}_{\text{I},{k}}}[\ell]\mathbf{C}_{\eta}^{-1/2,*}$\\
Vector of \ac{AWGN} at subcarrier $\ell$ & $\mathbf{n}[\ell]$\\
 $k$-th \ac{BS} equalizer filter at subcarrier $\ell$ &${\mathbf{G}_{}}_{k}^{}[\ell]$\\
Vector of $N_{\text{s},k}[\ell]$ estimated  symbols at subcarrier $\ell$& $\hat{\mathbf{s}}^{\text{UL}}_k[\ell]$\\
\hline
\end{tabular}
\end{table}

Once the uplink \ac{MSE} corresponding to all the $K$ users for a particular subcarrier has been determined, we can similarly define $\text{MSE}^{\text{UL}} = \sum^L_{\ell=1}\text{MSE}^{\text{UL}}[\ell]=\sum^K_{k=1}\sum^L_{\ell=1}\text{MSE}^{\text{UL}}_k[\ell]$ which is the overall system \ac{MSE} in the dual uplink when considering the symbols transmitted by all the users over all the subcarriers. This uplink \ac{MSE} can be represented in a compact form as follows
\begin{align}\label{fop}
\notag\text{MSE}^{\text{UL}}=& \text{tr}\Bigg[\Bigg(\mathbf{I}_{{N}_{\text{s}}L}+ \mathbf{T}^*\mathbf{C}_{}^{-1/2}\mathbf{\hat{H}}_{\text{e}} \Bigg( %
\mathbf{I}_{L}\otimes\mathbf{I}_{N_{\text{t}}}+(1+N)\;
\\
&\times\frac{L}{P_{\text{T}}}\parallel \mathbf{T}\parallel_{\operatorname{F}}^{{2}} \mathbf{I}_{N_{\text{t}}L}\Bigg)^{-1} \mathbf{\hat{H}}^*_{\text{e}} \mathbf{C}_{}^{-1/2,*}\mathbf{T}\Bigg)^{-1}\Bigg],
\end{align}
where $\mathbf{I}_{N_{\text{s}}L}$ is the $N_{\text{s}}L \times N_{\text{s}}L$ identity matrix and
\begin{align}
    \mathbf{\hat{H}}_{\mathbf{e}}={{\mathbf{\hat{H}}_{\text{B}}}}+\sum^{N_{}}_{n=1}\tilde{{\nu}}_n{{\mathbf{\hat{H}}}_{\text{c},{{n}}}},
\end{align}
where $\tilde{\boldsymbol{\nu}}$ is the main diagonal of the resulting matrix after applying $\tilde{\boldsymbol{\Theta}}=\mathbf{I}_{L}\otimes\boldsymbol{\Theta}$ and
\begin{align}
\notag
&\mathbf{T}^* =\text{blkdiag}\left(\mathbf{T}^*[1],\ldots,\mathbf{T}^*[L]\right), \mathbf{C}=\mathbf{I}_{L}\otimes \mathbf{C}_{\eta},\\\notag
&\mathbf{\hat{H}}^*_{\text{B}}=\text{blkdiag}\left(\mathbf{{\hat{H}}}^*_{\text{B}}[1],\ldots,\mathbf{\hat{H}}^*_{\text{B}}[L]\right),\\\notag
&\mathbf{E}^*_{\text{B}}=\text{blkdiag}\left(\mathbf{E}^*_{\text{B}}[1],\ldots,\mathbf{E}^*_{\text{B}}[L]\right),\\\notag
&{\mathbf{E}^*_{\text{c},n}}=\text{blkdiag}\left({\mathbf{E}^*_{\text{c},n}}[1],\ldots,{\mathbf{E}^*_{\text{c},n}}[L]\right),\forall n=1,\ldots ,N\notag,\\
&
\notag
    \notag {{\mathbf{\hat{H}}}^*_{\text{c},{{n}}}}\hspace{-0.0mm}=\hspace{-0.0mm}\text{blkdiag}\left(\hspace{-0.0mm} {{\mathbf{\hat{H}}}^*_{\text{c},{{n}}}}[1],\ldots,{{\mathbf{\hat{H}}}^*_{\text{c},{{n}}}}[L]\hspace{0mm}\right),\forall n=1,\ldots ,N\notag
\end{align}
are block diagonal matrices that stack all the precoders, the noise covariance matrices and the channel responses corresponding to the $K$ users at the $L$ subcarriers, such that $\mathbf{T}\in\mathbb{C}^{N_{\text{r}}KL \times  N_{\text{s}}}$, ${{\mathbf{\hat{H}}}^*_{\text{c},{{n}}}}\hspace{-1.2mm}\in\mathbb{C}^{N_{\text{r}}KL \times N_{\text{t}}L }$, $\mathbf{H}^*_{\text{B}}\in\mathbb{C}^{N_{\text{r}}KL \times N_{\text{t}}L }$, $\mathbf{E}^*_{\text{B}}\in\mathbb{C}^{N_{\text{r}}KL \times N_{\text{t}}L }$ and ${\mathbf{E}^*_{\text{c},n}}\in\mathbb{C}^{N_{\text{r}}KL \times N_{\text{t}}L }$. 

Using the previous compact notation, the \ac{MMSE} optimization problem for the virtual uplink can be formulated as follows
\begin{align}\label{op}
&\underset{\mathbf{\Theta}^{},\mathbf{T}_k[\ell],\forall k,\ell}{\text{arg}\;\text{min}}\;\;\text{MSE}^{\text{UL}}(\mathbf{T,\mathbf{\Theta}})
\\&
\text{s.t.}\;\sum_{k=1}^{K} \text{tr}({\mathbf{T}_{k}}[\ell]{\mathbf{T}^{*}_{k}}[\ell])\leq P_{\text{T}}[\ell],\forall \ell,\; \mathbf{\Theta}^{}\in\mathcal{D} \notag
,\;\notag
\end{align}
with $\text{MSE}^{\text{UL}}$ given in \eqref{fop}.
Note that after obtaining the precoders $\mathbf{T}_k[\ell]$ $\forall k,\ell$ and the receiving filters ${\mathbf{G}_{\text{MMSE},k}} [\ell]$ $\forall k,\ell$ in the dual uplink, we readily determine $\mathbf{P}_k[\ell]$ $\forall k,\ell$ and ${\mathbf{W}_{\text{MMSE},k}}[\ell]$ $\forall k,\ell$ in the downlink by means of \eqref{dualp} and the MMSE expression in \eqref{mmsefil}, respectively. 

 \subsection{Alternating \ac{MSE} minimization algorithm}\label{explal}
In this subsection, we develop an alternating algorithm to solve the \ac{MSE} minimization problem \eqref{op}. The frequency-flat \ac{IRS} phase-shift matrix $\mathbf{\Theta}$ and the frequency-dependent filters/precoders in the uplink and their downlink counterparts are alternately calculated until the \ac{MSE} reduction is not higher than a threshold $\delta$ or until a maximum number of iterations $\epsilon$ is reached. 

\Cref{PG} summarizes the steps of the proposed alternating minimization approach. \Cref{PG} starts determining the filters and downlink precoders, which are next used to determine the filters and precoders in the dual uplink through \eqref{dualwu} and \eqref{mmsefuk}. Then, the \ac{IRS} phase shift matrix $\mathbf{\Theta}=\text{diag}(\boldsymbol{\nu})$ is determined with the following iterative projected gradient algorithm
 \begin{equation}\label{algoritmo}
     {\boldsymbol{\nu}}^{(i)} =d\left(\boldsymbol{\nu}^{\left(i-1 \right)}_{}- \mu_{\boldsymbol{\nu}}\,{\nabla_{\boldsymbol{\nu}}}\;\text{MSE}^{\text{UL}}(\mathbf{T}^{(i)},\boldsymbol{\nu}^{(i)})\right).
 \end{equation}
The operator $d(\cdot)$ is a projector which enforces ${\nu}_n,\forall n$ to be a unitary modulus element and, thus, $\mathbf{\Theta}$ to belong to the set of feasible solutions $\mathcal{D}$, i.e., $d(\cdot)$ ensures that the \ac{IRS} matrix is diagonal with unit magnitude entries. 
Next, the \ac{MAC}-\ac{BC} duality expression in \eqref{dualp} and the \ac{MMSE} filter expression in \eqref{mmsefil} are used to compute the precoders and filters in the downlink, i.e., $\mathbf{P}_k[\ell]$ $\forall k,\ell$ and ${\mathbf{W}_{\text{MMSE},k}}[\ell]$ $\forall k,\ell$, respectively. By invoking the \ac{MAC}-\ac{BC} duality in both directions, this alternating procedure is repeated until achieving the stopping criterion in line 21.

The initial \ac{IRS} phase-shift matrix $\mathbf{\Theta}^{(0)}$ is set to a diagonal matrix whose non-zero entries have unit magnitude and a random phase from the interval $[0, 2\pi)$ (line 3). The initial block diagonal precoding matrix in the downlink $\mathbf{P}^{(0)}$ is constructed with the \ac{MRT} precoders for each user at each subcarrier (line 5) assuming the uniform power allocation ${P_{\text{T}}}_k[\ell]=\frac{P_{\text{T}}}{KL}$ $\forall k,\ell$. Notice that \Cref{PG} also updates at each iteration the power allocation per user and per subcarrier.

\vspace*{0.2cm}
Considering $$\mathbf{B}(\boldsymbol{\nu})\hspace{-1.0mm}=\hspace{-1.0mm}\mathbf{I}_{N_{\text{s}}L}+\mathbf{T}^*\mathbf{C}^{-1/2}\mathbf{H}_{\text{e}}\mathbf{H}^*_{\text{e}}\mathbf{C}^{-1/2,*}\mathbf{T}\frac{1}{1+(N+1)\frac{L}{P_{\text{T}}}{\parallel \mathbf{T} \parallel}_{\operatorname{F}}^{2}}$$ and $\text{MSE}^{\text{UL}}(\mathbf{T},\boldsymbol{\nu})=\text{tr}[\mathbf{B}^{-1}(\boldsymbol{\nu})]$, we get  the gradient  ${\nabla_{{\nu}_n}}\;\text{MSE}^{\text{UL}}(\mathbf{T,\boldsymbol{\nu}})$ used in \eqref{algoritmo}

\begin{align}\label{grt}
&\notag {\nabla_{{\nu}_n}}\text{MSE}^{\text{UL}}(\mathbf{T,{\nu}}) \hspace{-1mm}=\hspace{-1mm}-\text{tr}\left( \mathbf{B}^{-2}(\boldsymbol{\nu})\mathbf{T}^*\mathbf{C}^{-1/2}\mathbf{\hat{H}}_{c,n}\mathbf{\hat{H}}^*_{\text{e}}\mathbf{C}^{-1/2,*}\mathbf{T}\right)\\
&\times \frac{1}{1+(1+N)\frac{L}{P_{\text{T}}}{\parallel \mathbf{T}\parallel}^{2}_{\operatorname{F}}},
\end{align}
where we have used the equality $\boldsymbol{\nabla}_{\beta}\text{tr}(\mathbf{A}^{-1}(\beta))=-\text{tr} (\mathbf{A}^{-2}(\beta))\frac{\partial \mathbf{A}(\beta)}{\partial\beta}.$

\begin{algorithm}[t]
	\caption{Alternating \ac{MSE} Minimization PG}\label{PG}
\textbf{Input:} ${{\mathbf{C}_{\eta_k}}, \forall k, P_{\text{T}}, ~\mu_1,\delta,\epsilon, \mathbf{\hat{H}}}^*_{\text{B}}\in \mathbb{C}^{LN_{\text{r}}K\times LN_{\text{t}}}, {{\mathbf{\hat{H}}}^*_{\text{c},{{n}}}}\hspace{-0.5mm} \in \mathbb{C}^{LN_{\text{r}}K\times LN_{\text{t}}}, \forall n %
$
	\begin{algorithmic}[1]
		\State  \textbf{Initialize:} $i \gets 0$
\State $\theta_{n} \in \mathcal [0, 2\pi),\;\forall\;n$		
		\State $\mathbf{\Theta}^{(0)}=\text{diag}( e^{\operatorname{j}\theta_1},\ldots,e^{\operatorname{j}\theta_{N}}),\; \forall\;n$
\For	{$\ell=1:L$}	
		\State  $\hspace{-0.5mm}\mathbf{P}^{(0)}[\ell]=\Big[\mathbf{P}^{(0)}_1[\ell],\ldots,\mathbf{P}^{(0)}_K[\ell]\Big] \gets$  \ac{MRT} precoders
		\State $\hspace{-2mm}\mathbf{W}^{(0)}_{\text{MMSE}}[\ell]\hspace{-1.3mm}=\hspace{-1.2mm}\text{blkdiag}\hspace{-0.5mm}\Big(\hspace{-1.0mm}{\mathbf{W}^{(0)}_{\text{MMSE},1}}\hspace{-0.5mm}[\ell],\ldots,\hspace{-1mm}{\mathbf{W}^{(0)}_{\text{MMSE},K}}\hspace{-0.5mm}[\ell]\hspace{-1.0mm}\Big) $ $\hspace{-1.3mm}\gets$ \hspace{-1.3 mm}\eqref{mmsefil}
		\State $\mathbf{T}^{(0)}[\ell]=\text{blkdiag}\Big(\mathbf{T}^{(0)}_1[\ell],\ldots,\mathbf{T}^{(0)}_K[\ell]\Big) $ $\gets$  \eqref{dualwu}
		\State  $\mathbf{G}^{(0)}_{\text{MMSE}}[\ell]=\Big[{\mathbf{G}^{*(0)}_{\text{MMSE},1}}[\ell],\ldots,{\mathbf{G}^{*(0)}_{\text{MMSE},K}}[\ell]\Big]^* $ $\gets$ \eqref{mmsefuk}
\EndFor		
		\State $\mu_{\mathbf{\Theta}} \gets \mu_{1}$
		\Repeat

			\State $i \gets i+1$ 
			\vspace*{0.02cm}

\For	{$\ell=1:L$}	
		\State \hspace{-1mm}$\mathbf{P}^{(i)}[\ell]\hspace{-1mm}=\hspace{-1mm}\Big[\mathbf{P}^{(i)}_1[\ell],\ldots,\mathbf{P}^{(i)}_K[\ell]\Big]\hspace{-1mm} \gets$  \hspace{-1mm} \eqref{dualp} and $\mathbf{G}^{(i-1)}_k[\ell]$
	\State \small{$\hspace{-0.7mm}\mathbf{W}^{(i)}_{\text{MMSE}}[\ell]\hspace{-0.8mm}=\hspace{-0.8mm}\text{blkdiag}\hspace{-0.6mm}\Big(\hspace{-0.6mm}{\mathbf{W}^{(i)}_{\text{MMSE},1}}\hspace{-0.7mm}[\ell],\ldots,\hspace{-1.3mm}{\mathbf{W}^{(i)}_{\text{MMSE},K}}\hspace{-0.5mm}[\ell]\hspace{-1.0mm}\Big) $ $\hspace{-1.3mm}\gets$ \hspace{-1.3 mm}\eqref{mmsefil}}
		\State $\mathbf{T}^{(i)}[\ell]=\text{blkdiag}\Big(\mathbf{T}^{(i)}_1[\ell],\ldots,\mathbf{T}^{(i)}_K[\ell]\Big) \gets$  \eqref{dualwu}
		\State  $\mathbf{G}^{(i)}_{\text{MMSE}}[\ell]=\Big[{\mathbf{G}^{*(i)}_{\text{MMSE},1}}[\ell],\ldots,{\mathbf{G}^{*(i)}_{\text{MMSE},K}}[\ell]\Big]^*   \hspace*{-2mm} \hspace*{-1mm}\gets$ \eqref{mmsefuk}
\EndFor		

		\State $    {\boldsymbol{\nu}}^{(i)} =d(\boldsymbol{\nu}^{\left(i-1 \right)}- \mu_{\boldsymbol{\nu}}\,{\nabla_{\boldsymbol{\nu}}}\;\text{MSE}^{\text{UL}}(\mathbf{T^{\left(i\right)},\boldsymbol{\nu}^{\left(i\right)}}))$

		\While {{\hspace{-0.2mm}$\text{MSE}^{\text{UL}}\hspace{-0.2mm} \left(\mathbf{T}^{\left(i-1 \right)},\boldsymbol{\nu}^{\left(i-1 \right)}\right) \hspace{-0.5mm}\leq \text{MSE}^{\text{UL}}\left(\mathbf{T}^{\left(i\right)},\boldsymbol{\nu}^{\left(i \right)}\right)$}}
		
		\State $\mu_{\boldsymbol{\nu}} \gets \mu_{\boldsymbol{\nu}}/2 $
		\State $    {\boldsymbol{\nu}}^{(i)} =d(\boldsymbol{\nu}^{\left(i-1 \right)}_{}- \mu_{\boldsymbol{\nu}}\,{\nabla_{\boldsymbol{\nu}}}\;\text{MSE}^{\text{UL}}(\mathbf{T^{\left(i\right)},{\nu}^{\left(i\right)}}))$
		
		\EndWhile
		
		\Until \vspace{-6mm} \hspace{-0.5mm}
		\small{\begin{align*} \text{MSE}^{\text{UL}} \hspace{-0.5mm}&\left(\mathbf{T}^{\left(i-1 \right)},\boldsymbol{\nu}^{\left(i-1 \right)}\right) \hspace{-0.5mm} \\&- \text{MSE}^{\text{UL}}\hspace{-0.5mm} \left(\mathbf{T}^{\left(i \right)},\boldsymbol{\nu}^{\left(i \right)}\right)\hspace{-0.5mm}<\delta \;\,\text{or}\;~i\geq\epsilon\end{align*}}
		\end{algorithmic}
\textbf{Output:}\; $\mathbf{P}_k[\ell]$, ${\mathbf{W}_{\text{MMSE},k}}[\ell], \forall k,\ell$, $\mathbf{\Theta}\in \mathcal{D}$
\end{algorithm}

\section{Convergence Analysis}\label{Section_conv}

In this section, we conduct an analysis of the convergence properties of the proposed alternating MSE minimization PG algorithm. Algorithm 1 iterates over the precoders/filters (in both the \ac{BC} and the dual \ac{MAC} system) for the $L$ subcarriers, as well as on the diagonal elements of the frequency-flat \acs{IRS} phase-shift matrix. At each iteration, the algorithm produces $\mathbf{P}^{(i)}[\ell]$, $\mathbf{W}^{(i)}_{\text{MMSE}}[\ell]$, $\mathbf{T}^{(i)}[\ell]$, $\mathbf{G}^{(i)}_{\text{MMSE}}[\ell]$ and $\boldsymbol{\nu}^{(i)}$.

The precoders/filters for each subcarrier with $\ell=1,\ldots,L$ in both the BC and the dual MAC system are obtained using the MSE dualities as explained in \Cref{explal}. Initially, the precoders $\mathbf{P}^{(0)}[\ell]$ in the \ac{BC} are obtained via \ac{MRT} precoding, while the closed-form expression in equation \eqref{mmsefil} is used to compute the MMSE filters $\mathbf{W}^{(0)}_{\text{MMSE}}[\ell]$ in the BC. Next, the precoders $\mathbf{T}^{(0)}[\ell]$ in the dual MAC system are computed using the BC-MAC duality given by \eqref{dualwu}, and the closed-form expression in \eqref{mmsefuk} is employed to obtain the MAC MMSE filters $\mathbf{G}^{(0)}_{\text{MMSE}}[\ell]$.
After these initialization operations, during the $i$-th iteration, the  precoders $\mathbf{P}^{(i)}[\ell]$ are computed using the MAC-BC duality as expressed in \eqref{dualp}, and the filters $\mathbf{W}^{(i)}_{\text{MMSE}}[\ell]$ are obtained through the closed-form expression in (21). Additionally, the MAC precoders  $\mathbf{T}^{(i)}[\ell],\forall \ell$ and the MAC filters $\mathbf{G}^{(i)}_{\text{MMSE}}[\ell],\forall \ell$ are once again obtained using the BC-MAC duality in \eqref{dualwu} and the closed-form expression in \eqref{mmsefuk}, respectively.
The \ac{IRS} phase shift matrix $\mathbf{\Theta}=\text{diag}(\boldsymbol{\nu})$ is determined by using the gradient ${\nabla_{\boldsymbol{\nu}}}$ $\text{MSE}^{\text{UL}} (\mathbf{T}^{},\boldsymbol{\nu}^{})$ as described in \eqref{algoritmo}.

  To prove convergence, it is sufficient to ensure that the performance metric $\text{MSE}^{\text{UL}} (\mathbf{T}^{\left(i \right)},\boldsymbol{\nu}^{\left(i \right)})$ either decreases or remains constant at each iteration. Since the performance metric is lower bounded, this property guarantees that the proposed algorithm converges to a local optimum. Hence, we
need to verify that each sequence $\lbrace \text{MSE}^{\text{UL}}({{\mathbf{T}}},\underline{\boldsymbol{\nu}}^{(i)})\rbrace$ and $\lbrace \text{MSE}^{\text{UL}}(\underline{{\mathbf{T}}}^{(i)},\boldsymbol{\nu}^{(i)})\rbrace$ converges to a local minimum, where $\underline{\B{X}}$ denotes the fixed variables.

\subsubsection {Convergence of the sequences $\lbrace \text{MSE}^{\text{UL}}({{\mathbf{T}}},\underline{\boldsymbol{\nu}}^{(i)})\rbrace$}

The convergence of the MSE duality procedure in conventional communication systems without IRSs (or a fixed IRS) has been previously analyzed. In \cite{4355332} (please, see Sections III and IV),
the authors prove monotonic convergence, and it is observed that the algorithm converges rapidly in the initial iterations. However, proving global optimality is challenging. In this case, for Algorithm 1, the sequences produced $\lbrace \text{MSE}^{\text{UL}}({\mathbf{T}}^{(i)},\underline{\boldsymbol{\nu}}^{(i)})\rbrace$ converge to a local minimum. 
However, this assumptions is based on a fixed $\mathbf{\Theta}=\text{diag}(\boldsymbol{\nu})$. In our implementation, we update $\boldsymbol{\nu}^{(i)}$ for the $i$-th iteration within expressions \eqref{mmsefil}  and \eqref{mmsefuk} in ${\mathbf{\hat{H}}_\text{e}}[\ell]$ (cf. \eqref{heqv}). As a result, it becomes crucial to ensure the convergence of the sequences $\lbrace \text{MSE}^{\text{UL}}(\underline{{\mathbf{T}}}^{(i)},\boldsymbol{\nu}^{(i)})\rbrace$ for a given or fixed precoder $\mathbf{T}$.

\subsubsection{ Convergence of the sequences $\lbrace \text{MSE}^{\text{UL}}(\underline{{\mathbf{T}}}^{(i)},\boldsymbol{\tilde{\nu}}^{(i)})\rbrace$}
Let us define the auxiliary \ac{IRS} phase-shift matrix as follows  $$\tilde{\boldsymbol{\nu}}^{ \left(i \right)}=\tilde{\boldsymbol{\nu}}^{\left(i-1 \right)}_{}- {\mu}_{{\boldsymbol{\nu}}}\,\nabla_{{\boldsymbol{\nu}}}\;\text{MSE}^{\text{UL}}({\mathbf{T}^{(i)}},\tilde{\boldsymbol{\nu}}^{(i)})$$ at the $\ell$-th iteration, where  $\tilde{\boldsymbol{\nu}}^{}$ represents the unconstrained solution for the optimization problem in \eqref{op}. Note that ${{\mathbf{T}}}$ and ${\tilde{\boldsymbol{\nu}}}$ are obtained in steps 15 and 17 of Algorithm 1, respectively, by neglecting the projection operator $d(\cdot)$ for the IRS design. In this case, the sum-MSE sequences produced by the alternating minimization algorithm, $\lbrace \text{MSE}^{\text{UL}}({\mathbf{T}}^{(i)},\underline{\mathbf{\tilde{\boldsymbol{\nu}}}})\rbrace$ and $\lbrace \text{MSE}^{\text{UL}}(\underline{{\mathbf{T}}},\tilde{\boldsymbol{\nu}}^{(i)})\rbrace$ converge to a local minimum with the proper adjustment of the step size  ${\mu}_{{\boldsymbol{\nu}}}$, and by using a line or Armijo’s search \cite{1101194}.  %
To obtain ${\boldsymbol{\Theta}}=\text{diag}(\boldsymbol{\nu})$ within the feasible set  $\mathcal{D}$, we employ the projector $d(\cdot)$, leading to the sequence  $\lbrace \text{MSE}^{\text{UL}}(\underline{{\mathbf{T}}},\boldsymbol{\nu}^{(i)})\rbrace$. %
Hence, we need to verify that this sequence with constrained solutions converges to a local minimum, as we do next.

\subsubsection {Convergence of the sequences $\lbrace \text{MSE}^{\text{UL}}(\underline{{\mathbf{T}}}^{(i)},\boldsymbol{\nu}^{(i)})\rbrace$}

The projector $d(\cdot)$ is employed to map $\boldsymbol{\tilde{\nu}}$ and obtain $\mathbf{\Theta}$ onto the non-convex set $\mathcal{D}$, i.e., $\boldsymbol{\Theta}=\text{diag}(\boldsymbol{{\nu}})$. Note that the unit modulus constraint defines a Riemannian manifold (cf. \cite{absil2009optimization} and \cite{lee2013smooth}), which is intrinsically non-convex. Nevertheless, it has been proven in \cite{9367432} that if the function $\text{MSE}^{\text{UL}}(\mathbf{T},\boldsymbol{\nu})$ is differentiable with a continuous gradient, the convergence of $\lbrace \text{MSE}^{\text{UL}}(\underline{{\mathbf{T}}},\boldsymbol{\nu}^{(i)})\rbrace$ to $(\underline{{\mathbf{T}}},\boldsymbol{\nu}^{\text{(o)}})$ is ensured because $\mathcal{D}$ is a compact set (see  \cite[Theorem 1]{9367432} and \cite[Appendix]{yang2019inexact}).

Finally, it is essential to highlight that our approach based on alternating minimization over the individual variables provides a $q$-linear convergence rate over both $\mathbf{T}$ and $\boldsymbol{\nu}$ under practical assumptions. Let $(\mathbf{T}^{\text{(o)}},\boldsymbol{\nu}^{\text{(o)}})$ be the local minimum of $\text{MSE}^{\text{UL}}(\mathbf{T},\boldsymbol{\nu})$ on a neighborhood $\mathcal{N}$ of $(\mathbf{T}^{\text{(o)}},\boldsymbol{\nu}^{\text{(o)}})$ over which the function $\text{MSE}^{\text{UL}}(\cdot)$ is convex. Given that $\text{MSE}^{\text{UL}}(\mathbf{T},\boldsymbol{\nu})$ is a smooth function (i.e., continuous and differentiable) over such $\mathcal{N}$, and taking into account that the Hessian matrix of $\text{MSE}^{\text{UL}}(\cdot)$ is positive definite, alternating minimization over each variable ensures a $q$-linear convergence rate to  $(\mathbf{T}^{\text{(o)}},\boldsymbol{\nu}^{\text{(o)}})$ \cite{bezdek2003convergence}. This condition holds when all the available channel spatial degrees of freedom are utilized, i.e., for mid-to-high \ac{SNR} levels. This regime is practical for some \ac{IRS}-aided scenarios.
The empirical verification of these findings is presented in the numerical results.

\section{Computational Complexity Analysis}\label{Section V}
In this section, we analyze the computational complexity of \Cref{PG}. The complexity orders for the main steps are summarized in \Cref{tablecomlejidad}. As shown, the complexity of step 5 (computation of \ac{MRT} precoders) is bounded by $\mathcal{O}\left( N_{\text{r}}{N_{\text{t}}}^2KL\right)$ since $KL$ \acp{SVD} are necessary. In step 17 (update of $\boldsymbol{\nu}^{(i)}$), the gradient given by \eqref{grt} is used and its computational complexity order is bounded by $\mathcal{O}\left( LN_{\text{r}}N_{\text{t}}KN\epsilon\right)$. Note that this line is performed $\epsilon$ times at most. The complexity order of lines 18 and 20 used to compute the \ac{MSE} performance metric \eqref{fop} is $\mathcal{O}\left( LN_{\text{r}}N_{\text{t}}KN\epsilon\right)$. We conclude that \Cref{PG} leads to an overall complexity order of $\mathcal{O}\left( N_{\text{r}}{N_{\text{t}}}^2KL\right)$. 
\begin{table}[h!]
\vspace{-2mm}
	\centering
	\caption{{Computational complexity of \Cref{PG}}.}
   \vspace{-2mm}
	\setlength{\tabcolsep}{4pt}
	\def\arraystretch{1.1}
	\label{tablecomlejidad}
	\begin{tabular}{lllll}
	\hline

\multicolumn{1}{c}{\begin{tabular}[c]{@{}c@{}}  \end{tabular}}                   & \multicolumn{1}{c}{\textbf{Operation}}   &\multicolumn{1}{c}{\begin{tabular}[c]{@{}c@{}}\textbf{Num. of} \textbf{operations}\end{tabular}}                & \multicolumn{1}{c}{\textbf{Complexity order}}

		\\ \cline{2-5}  
	\hline
	
		\multicolumn{1}{c}{\begin{tabular}[c]{@{}c@{}}  \end{tabular}}                   & \multicolumn{1}{c}{\begin{tabular}[c]{@{}c@{}}Compute MRT Precoders\\  (step 5)\end{tabular}}                  & \multicolumn{1}{c}{1} & \multicolumn{1}{c}{$\mathcal{O}\left( N_{\text{r}}{N_{\text{t}}}^2KL\right)$}

		\\  \cline{2-5}
		\multicolumn{1}{c}{}                   & \multicolumn{1}{c}{\begin{tabular}[c]{@{}c@{}}Compute $\boldsymbol{\nu}^{(i)}$  (step 17)\end{tabular}}&                  \multicolumn{1}{c}{$\epsilon$} & \multicolumn{1}{c}{$\mathcal{O}\left(L N_{\text{r}}N_{\text{t}}KN\epsilon\right)$}                    
		\\ \cline{2-5} 
		
		\multicolumn{1}{c}{}                    &   \multicolumn{1}{c}{\begin{tabular}[c]{@{}c@{}}Overall MSE evaluation\\ (steps 18 and 20)\end{tabular}}& \multicolumn{1}{c}{$\epsilon$}
&  \multicolumn{1}{c}{$\mathcal{O}\left(L N_{\text{r}}N_{\text{t}}KN\epsilon\right)$}                        & \\\hline

		\multicolumn{1}{c}{}                    &   \multicolumn{1}{c}{\textbf{Overall}}      &                 & \multicolumn{1}{c}{$\mathcal{O}\left( N_{\text{r}}{N_{\text{t}}}^2KL\right)$}                        &                                          \\\hline
	\end{tabular}
\end{table}

\section{Simulation Results}\label{Section VI}

In this section, we present the results of computer simulations carried out to assess the performance of the wideband \ac{IRS}-aided \ac{MU} \ac{MIMO} system designed with the alternating \ac{MSE} minimization algorithm proposed in \Cref{Section IV}. We consider a wideband \ac{IRS}-aided \ac{MU} \ac{MIMO} system with $K=3$ users having $N_{\text{r}}=4$ antennas each, $L=32$ subcarriers, a \ac{BS} with $N_{\text{t}}=9$ antennas which allocates 2 data streams per user at subcarrier $\ell$, i.e., $N_{\text{s},k}[\ell]=2$ $\forall k,\ell$, and an \ac{IRS} with different numbers of elements $N\in \lbrace 9, 16, 25, 36, 49, 64, 81, 100\rbrace$. 
Although some authors assume that the estimation error level at the receivers is lower or even negligible, our approach is more general and can be adapted to those cases.
The parameters of the \ac{mmWave} channel model are set to $L_{\text{D}}$ = 8 delay taps, ${N_{\text{path}}}_{\text{B}}$ channel paths for the \ac{BS}-\ac{IRS} links and ${N_{\text{path}}}_{\text{I}}$ paths for the \ac{IRS}-users links, respectively. The \acp{AoA} and the \acp{AoD} are assumed to be random and uniformly distributed over the interval $[0, \pi]$ as in \cite{8606454}. The relative delays $\tau_{n}$ are also random and uniformly distributed over the range $\tau_{n} \in [0, (L_{\text{D}}-1)T_{\text{s}}]$, with $T_{s} = 1/f_{s}$ and $f_{s} = 1760$ MHz. The complex-valued channel gains are i.i.d. standard Gaussian random variables, i.e., $\beta_{m} \sim \mathcal{N}_{\mathbb{C}}(0, 1)$. The central carrier frequency is assumed to be $f_{\text{c}} = 28$ GHz and the signal bandwidth is set to $400 \;\text{MHz}$. 

All the reported results are obtained after averaging $C_{\text{R}}=1000$ channel realizations. Performance is evaluated in terms of the downlink achievable sum-rate \eqref{RW} and the \ac{MSE} between the transmitted and estimated symbols. We assume that the noise covariance matrix has the form $\mathbf{C}_{\eta_k}=\sigma^2_{{\eta}_k} \mathbf{I}$, with equal noise power for the noise during data transmission and training stages, i.e., $\sigma^2_{{\eta}_k} =\sigma^2_{{\eta}}=1$ $\forall k$. Therefore, the per subcarrier \acl{SNR} is given by $\text{SNR (dB)}=10 \;\text{log}_{10} (P_{\text{T}}/L)$. 
The channel responses of the direct and cascaded channels are assumed to be imperfectly known, following a stochastic model where $\mathbf{E}_{\text{B},k}[\ell]$ and $\mathbf{E}_{\text{c},k,n}[\ell]$ are assumed to follow a complex Gaussian distribution  $\mathcal{N}_{\mathbb{C}}(\mathbf{0},\frac{L}{P_{\text{T}}}\mathbf{C}_{\eta_k})$, similar to the approaches in \cite{9999550,5711691}.
Note that this assumption entails a lower estimation error while the \ac{SNR} increases.
Finally, the maximum number of iterations $\epsilon$ in \Cref{PG} is set to 100.
\Cref{STsm} summarizes the system model configuration considered in the computer simulations. 
\begin{table}[h!]
\vspace{-2mm}
\centering
\caption{{Simulation parameter setting.}}\label{STsm}
  \vspace{-2mm}
	\setlength{\tabcolsep}{5pt}
	\def\arraystretch{1.15}
\begin{tabular}{r|l}
\hline
 {\textbf{Parameter}} & {\textbf{Setting}}  \\ \hline

Number of users & $K=3$\\
Antennas per users & $N_{\text{r}}=4$\\
Antennas at the \ac{BS} & $N_{\text{t}}=16$\\
Number of subcarriers & $L=32$\\
Number of data streams per user& $N_{\text{s},k}[\ell]=2,\forall k,\ell$\\
Number of \ac{IRS} elements & $N \in \lbrace 9 \sim 144 \rbrace$\\
BS-IRS channel propagation paths &$N_{\text{path}_{\text{B}}}=4$\\ 
IRS-users channel propagation paths &$N_{\text{path}_{\text{I}}}\in \{ 2,3,4\}$\\ 
Channel delay paths &$L_{\text{D}}=8$\\
Central carrier frequency & $f_{\text{c}}=$ 28 GHz\\ 
Covariance  of estimation error (downlink) & $\frac{L}{P_{\text{T}}}\mathbf{C}_{\eta_k},\forall k$ \\
Channel realizations & $C_{\text{R}}=1000$\\
Max. number of iterations (\Cref{PG}) & $\epsilon=100$\\
\hline
\end{tabular}
\end{table}

In the first simulation experiment, we compare the performance of the system designed according to \Cref{PG} with the following five baseline approaches:
\begin{enumerate}
\item The maximum achievable sum-capacity obtained by assuming \ac{DPC} and optimizing the \ac{IRS}. This approach constitutes an upper bound on the sum-rate and has been designed with the power iterative waterfilling algorithm proposed in \cite[Algorithm 2]{1412050}.
In this approach, we obtain the precoders for both the downlink and uplink systems using the iterative waterfilling algorithm presented in \cite{1412050}. Meanwhile, the \ac{IRS} phase shift matrix is updated at each iteration by following Algorithm 1. Note that this scheme yields the sum-capacity of the system for each equivalent channel $\mathbf{H}_{\text{e}}[\ell]$ provided by the IRS setup at each step. This is achieved by configuring the optimal precoders (optimal transmit policies) via \ac{DPC} under perfect CSI.

 \item \ac{AF O-Ps}. The precoders and the \ac{IRS} matrix are optimized through \Cref{PG} but the \ac{IRS} matrix is assumed to be unconstrained, i.e., the magnitude of the diagonal entries of the phase-shift matrix is not constrained to be $1$, and thus the \ac{IRS} is able to modify both the amplitude and the phase of the impinging signals. For the sake of fairness, we assume that $\|\mathbf{\Theta}\|_{\operatorname{F}}^2=N$. 
 This strategy is employed as an alternative upper bound for the design of the IRS phase shift matrix, given that achieving a globally optimal solution requires the use of the exhaustive search technique, which is computationally impractical, even for a system of moderate size. 
	\item \ac{R-IRS O-Ps}. The precoders are optimized via the \ac{MAC}-\ac{BC} and the \ac{BC}-\ac{MAC} dualities in \Cref{PG} while the phases of the diagonal \ac{IRS} phase-shift matrix are random uniform variables over $[0,2\pi]$, i.e., there is no control of the \ac{IRS} phase-shift matrix.
	\item \ac{O-IRS MRT-Ps}. The precoders are designed as \ac{MRT} precoders while the \ac{IRS} phase-shift matrix is optimized via \Cref{PG}.
	\item \ac{No-IRS MRT-Ps}. The precoders are designed as \ac{MRT} precoders and the \ac{IRS} is not deployed to assist the communication system, i.e., $\mathbf{\Theta}=\mathbf{0}$. Only the direct link between the \ac{BS} and the users is available. 
\end{enumerate}	
	In all the considered approaches (the proposed one and the five baselines), the wideband receiving filters $\mathbf{W}_k[\ell]$ $\forall \ell,k$ are designed according to the \ac{MMSE} criterion. It is also worth remarking that perfect \ac{CSI} is assumed in this first experiment. Note that the \ac{MMSE} expressions are readily adapted to perfect \ac{CSI} by setting the  error covariance matrix to $\mathbf{0}$.

\begin{figure}[h!]
\vspace{-2mm}
 \includegraphics[width=0.94\linewidth]{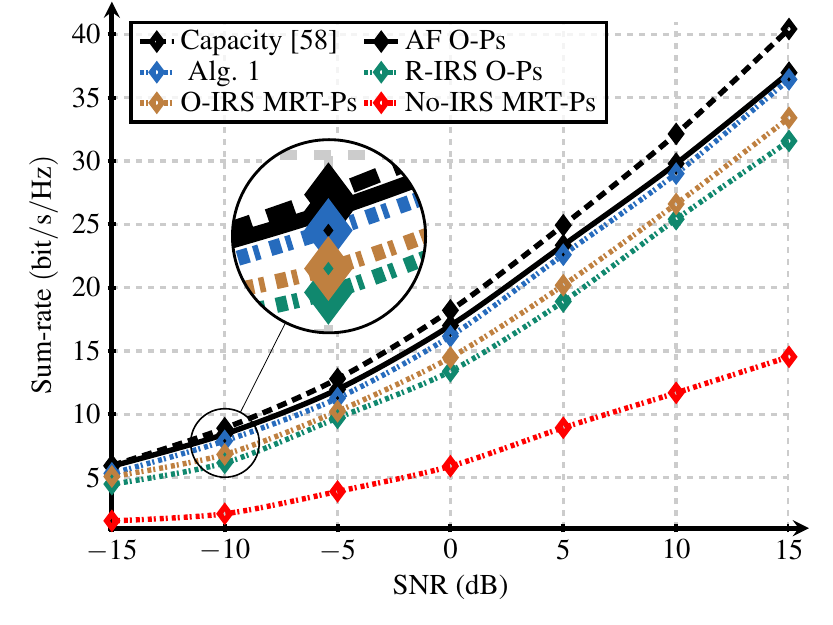}
 \vspace{-4.3mm}
 \caption{Sum-rate $(\text{bit}/\text{s}/\text{Hz})$ vs $\text{SNR\;(dB)}$ for $K=3$ users, $N_{\text{r}}=4$, $N_{\text{t}}=9$, $L=32$, $N=25$, and $N_{\text{s},k}[\ell]=2,\forall k,\ell$. }
  \label{fig3a}
\end{figure}

\Cref{fig3a} plots the achievable sum-rate obtained with the above mentioned approaches. As shown, \ac{AF O-Ps} provides the highest achievable sum-rate since the wideband precoders are optimized via \Cref{PG} and the \ac{IRS} phase-shift matrix entries have no magnitude constraint. The proposed \ac{PG}-based approach leads to a performance higher than that obtained with the \ac{R-IRS O-Ps} and the \ac{O-IRS MRT-Ps} baseline approaches. Recall that \ac{R-IRS O-Ps} does not optimize the \ac{IRS} phases while \ac{O-IRS MRT-Ps} does not account for the interference when designing the \ac{BS} wideband precoders. The \ac{No-IRS MRT-Ps} strategy provides the lowest system performance because only the direct channel is available. Thus, the system has a reduced capacity and the wideband precoder design does not take into account the \ac{MU} interference. It is also observed that the proposed \ac{PG} approach via \Cref{PG}
does not lead to a significant gap w.r.t. the unconstrained \ac{AF O-Ps} strategy and comes close to the maximum system capacity. Hence, the obtained results are very appealing because the proposed design algorithm is able to efficiently exploit the deployment of a passive IRS to improve the system performance.  

From \Cref{fig3a}, we can clearly observe a significant gain provided by the PG-based approach over the No-IRS MRT-Ps strategy, representing the improvement brought about by the IRS in enhancing the system performance. However, this gain is influenced by the quality of the channels between the IRS and the users, which is determined by the positioning of the IRS in practical deployment scenarios. Thus far we have considered $N_{{\text{path}}_\text{B}}=$$N_{{\text{path}}_\text{I}}=4$. In \Cref{tablagananciaIRS}, we define $G_{\text{IRS}} ~ (\text{bit}/\text{s}/\text{Hz})$ as the difference between the achievable sum-rate of the PG-based approach and the No-IRS MRT-Ps strategy. We also vary the number of reflection paths in the channels between the IRS and the users ($N_{{\text{path}}_\text{I}}$).  As expected, the results demonstrate that the gains $G_{\text{IRS}} ~ (\text{bit}/\text{s}/\text{Hz})$  decrease as 
the IRS-user links become weaker, indicating that the effectiveness of the IRS is closely related to the quality of these channels.

\begin{table}[htpb]
\vspace{-2mm}
\centering
\caption{\label{tablagananciaIRS}$G_{\text{IRS}} ~ (\text{bit}/\text{s}/\text{Hz})$ for different IRS-user channel conditions.}
  \vspace{-2mm}
  \def\arraystretch{1.1}
\begin{tabular}{ll|l|l|l}
\hline
 & \textbf{Num. of paths} & \textbf{5 dB}  & \textbf{10 dB} & \textbf{15 dB} \\ \hline
  & $N_{{\text{path}}_\text{I}}=2$ & 2.87 & 6.55  & 10.02 \\ 
 & $N_{{\text{path}}_\text{I}}=3$ &  10.75& 14.52 &  18.31\\ 
  & $N_{{\text{path}}_\text{I}}=4$ &  13.62&  17.30 &  21.90 \\ \hline
\end{tabular}
\end{table}

\begin{figure}[h!]
\centering
\vspace{-4.2mm}
 \includegraphics[width=0.93\linewidth]{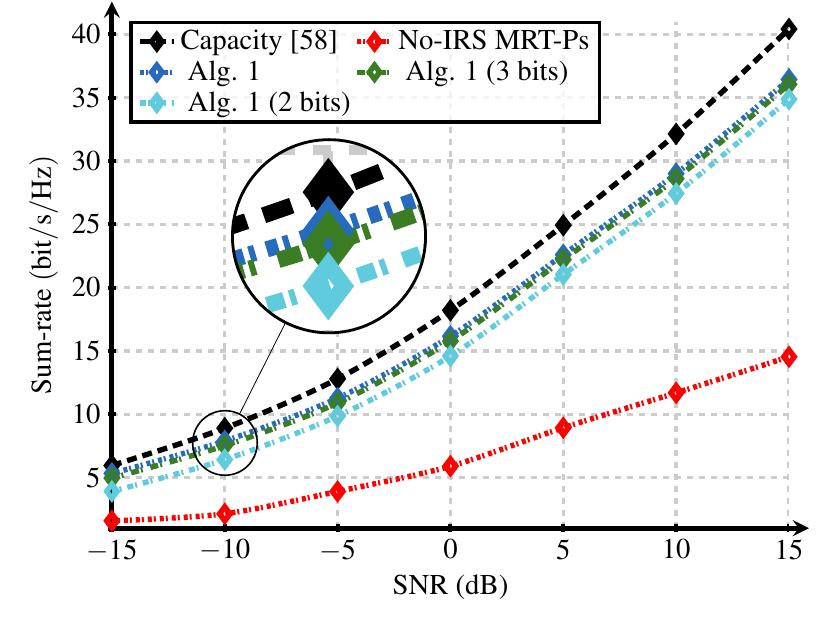}
  \vspace{-4.2mm}
 \caption{Sum-rate $(\text{bit}/\text{s}/\text{Hz})$ vs $\text{SNR\;(dB)}$ for $K=3$ users, $N_{\text{r}}=4$, $N_{\text{t}}=9$, $L=32$, $N=25$, $N_{\text{s},k}[\ell]=2,\forall k,\ell$ and discrete phase shifts (2 and 3 bits). }
  \label{fig3ar}
\end{figure}
{

We considered infinite resolution variable phase shifts to implement the IRS matrix. However, in practical scenarios, phase shifters are typically limited to a finite number of discrete values. To evaluate the impact of this quantization, we performed simulations for phase shift resolutions of 3 and 2 bits.
}
{The results in \Cref{fig3ar} show that the performance loss due to quantization is negligible when using \ac{IRS} phase shifting matrices with 8 available phases (3 bits). On the other hand, for \ac{IRS} phase shifting matrices with only 4 quantized values (2 bits), the performance loss is moderate.
}

\begin{figure}[h!]
\centering
\vspace{-1mm}
 \includegraphics[width=0.93\linewidth]{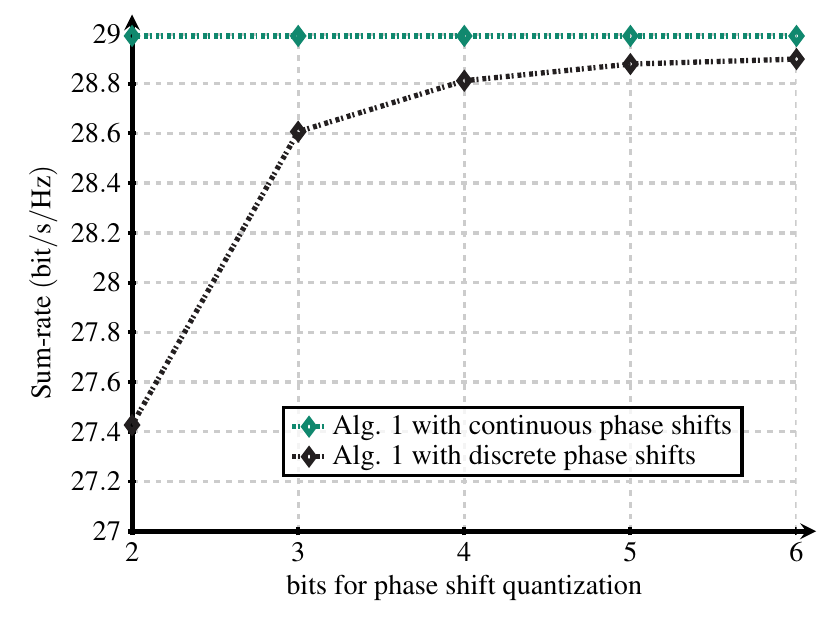}
  \vspace{-4.5mm}
 \caption{Sum-rate $(\text{bit}/\text{s}/\text{Hz})$ vs the number of bits used for phase-shift quantization levels for $K=3$ users, $N_{\text{r}}=4$, $N_{\text{t}}=9$, $L=32$, $N=25$,  $\text{SNR}=10\;\text{dB}$ and $N_{\text{s},k}[\ell]=2,\forall k,\ell$. }
  \label{figbits}
\end{figure}

In \Cref{figbits}, we present the sum-rate for a specific setup with $K=3$ users, $N_{\text{r}}=4$, $N_{\text{t}}=9$, $L=32$, $N=25$, $\text{SNR}=10\;(dB)$, and $N_{\text{s},k}[\ell]=2$ streams per user at subcarrier $\ell$. The continuous-valued IRS phase-shift matrix obtained via Algorithm 1 is used as the baseline for comparison. The figure shows that the sum-rate degradation due to discretizing the IRS phase shifts is lower than 2 $\text{bit}/\text{s}/\text{Hz}$, even when considering only 4 possible quantization phases (2 bits). Furthermore, this degradation effect steeply decreases when more bits are used for the phase-shift quantization.

\begin{figure}[h!]
\vspace{-3mm}
  \includegraphics[width=.93\linewidth]{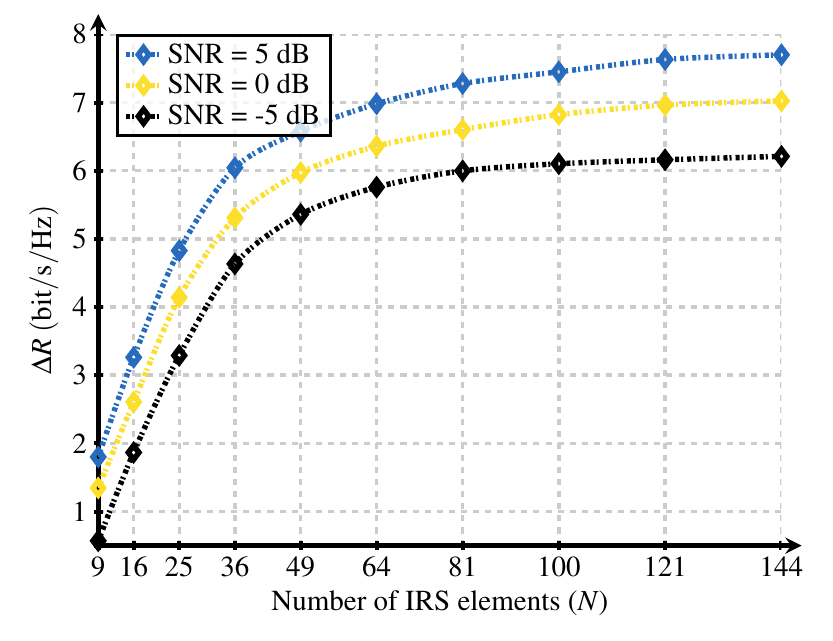}  
  \vspace{-4.3mm}
 \caption{Sum-rate increase $\Delta {R} ~ (\text{bit}/\text{s}/\text{Hz})$ vs $N$ for $K=3$ users, $N_{\text{r}}=4$, $N_{\text{t}}=9$, $\text{SNR}\;(\text{dB})\in{\lbrace -5, 0, 5\rbrace}$, $L=32$, $N\in{\lbrace9,16,25,36,49,64,81,100,121,144\rbrace}$, and $N_{\text{s},k}[\ell]=2,\forall k,\ell$.}
  \label{fig3c}
 
\end{figure}

In the next experiment, we quantify the performance improvement obtained when optimizing the \ac{IRS} phase-shift matrix with \Cref{PG}. More specifically, we measure the impact of the phase-shift optimization at the \ac{IRS} and evaluate the performance gains obtained when increasing the number of \ac{IRS} elements $N$. For that, we define the achievable sum-rate increase $\Delta {R} \;(\text{bit}/\text{s}/\text{Hz})$ as the difference between the achievable sum-rate obtained with \Cref{PG} and that obtained with \ac{R-IRS O-Ps}. The obtained results are presented in 
\Cref{fig3c}. %
As expected, the larger the size of the \ac{IRS} is, the higher the resulting gains are.  \Cref{fig3c} also shows how $\Delta {R} \;(\text{bit}/\text{s}/\text{Hz})$ increases with $\text{SNR\;(dB)}$. This effect is because of the optimization of the \ac{IRS} phase-shift matrix and, therefore, the control of the \ac{MU} interference has a larger impact on the system performance in the high \ac{SNR} regime.

 Another interesting observation from the results in the figure is the saturation of the performance gains $\Delta {R} (\text{bit}/\text{s}/\text{Hz})$ beyond a certain number of IRS elements. This behavior is inherently linked to the specific $N_{\text{r}}\times N_{\text{t}}$ MIMO wideband ($L$ subcarriers) configuration adopted for the \ac{MU} scheme with $K$ users, as well as the considered channel rank for the simulations. The system setup will determine how much improvement a large number of \ac{IRS} elements can provide to the communication system.

\begin{figure}[htpb]
\vspace{-2mm}
  \includegraphics[width=0.9\linewidth]{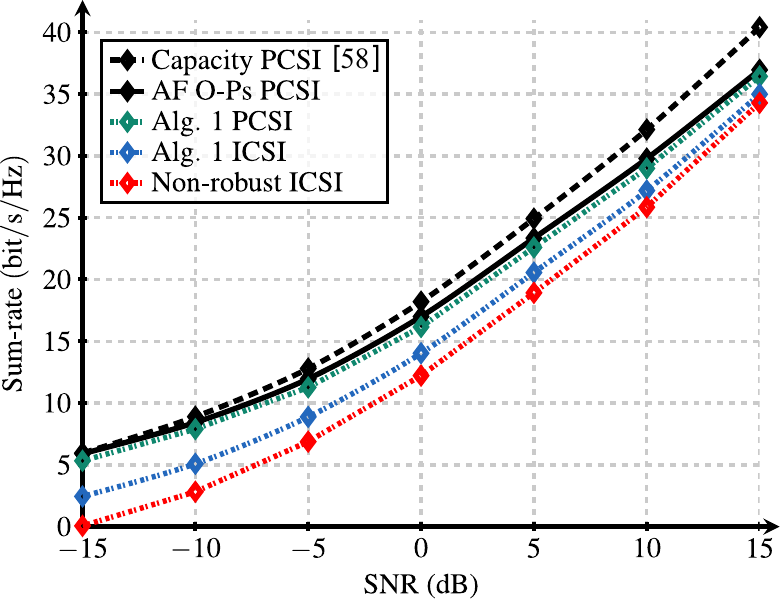}  
  \vspace{-2.5mm}
 \caption{Sum-rate $(\text{bit}/\text{s}/\text{Hz})$ vs $\text{SNR\;(dB)}$ for $K=3$ users, $N_{\text{r}}=4$, $N_{\text{t}}=9$, $N=25$, $L = 32$, imperfect {CSI}, and $N_{\text{s},k}[\ell]=2,\forall k,\ell$. }
  \label{fig4a}
\end{figure}

\begin{figure}[htpb]
\vspace{-6.0mm}
  \includegraphics[width=0.93\linewidth]{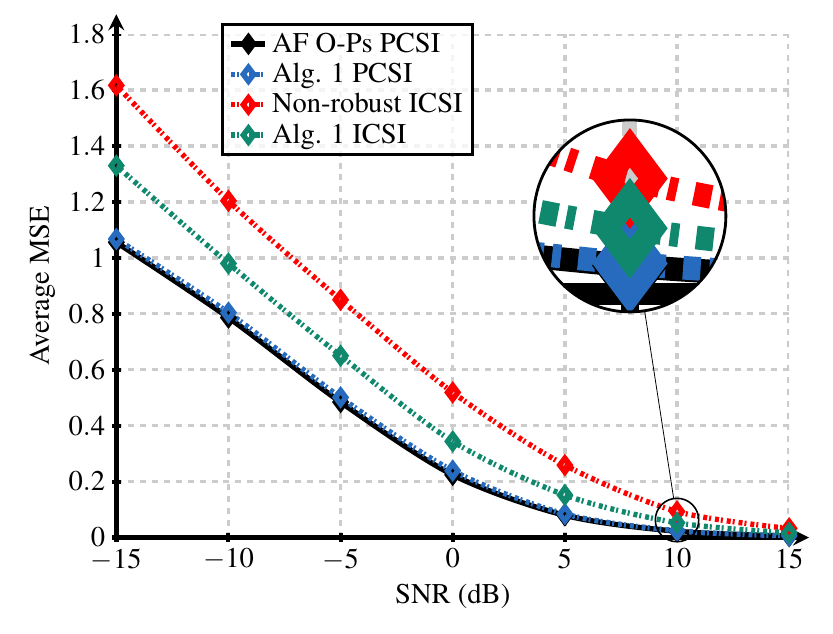}  
   \vspace{-4.6mm}
 \caption{Average MSE vs $\text{SNR\;(dB)}$ for $K=3$ users, $N_{\text{r}}=4$, $N_{\text{t}}=9$, $N=25$, $L = 32$, imperfect \ac{CSI}, and $N_{\text{s},k}[\ell]=2,\forall k,\ell$. }
  \label{fig4b}
\end{figure}

In the following, we evaluate (in terms of the achievable sum-rate) the impact of the \ac{CSI} estimation errors and compare the proposed solution with a baseline (labeled non-robust) that does not take into account the uncertainty introduced by the imperfect \ac{CSI} (Algorithm 1 in \cite{9547111}). We also include an approach where the precoders and the \ac{IRS} phase-shift matrix are determined with the proposed algorithm but assuming perfect \ac{CSI}. Finally, we included the sum-capacity and the \ac{AF O-Ps} solutions with perfect \ac{CSI} as benchmarks. 

As shown in \Cref{fig4a}, the non-robust strategy leads to the worst system performance since it does not consider imperfect \ac{CSI} and neglects the knowledge about the error statistics. This behavior is more apparent in the low \ac{SNR} regime, where the channel estimation errors are larger.
Conversely, \Cref{PG} considers the error statistics in the estimation of the direct and the cascaded channels, and outperforms the non-robust approach leading to a decreasing  gap w.r.t. the perfect \ac{CSI}. This decreasing  gap is observed since the estimation errors decrease in the high \ac{SNR} regime, and thus the system performance of the imperfect \ac{CSI} scheme approaches that of the perfect \ac{CSI} scenario.
 We remark that the behavior of the gap between the sum-rates achieved with the proposed solution in Algorithm 1 and with the non-robust strategy is influenced by the assumed covariance model for the errors, which leads to larger channel estimation errors in the low SNR regime. Nevertheless, our proposed solution in Algorithm 1 remains independent of the specific error model and the corresponding covariance matrix used for modeling the errors.

In \Cref{fig4b}, we compare the proposed and the non-robust approaches in terms of the \ac{MSE}. We also consider two benchmarks with perfect \ac{CSI}, namely the proposed approach and the \ac{AF O-Ps} scheme. It can be observed that the non-robust approach leads to the worst system performance, i.e., the highest \ac{MSE}, whereas the proposed solution for imperfect \ac{CSI} outperforms the non-robust approach (especially for low \ac{SNR} values), and comes close to the considered benchmarks.

\begin{figure}[h!]
\vspace{-3mm}
  \includegraphics[width=0.93\linewidth]{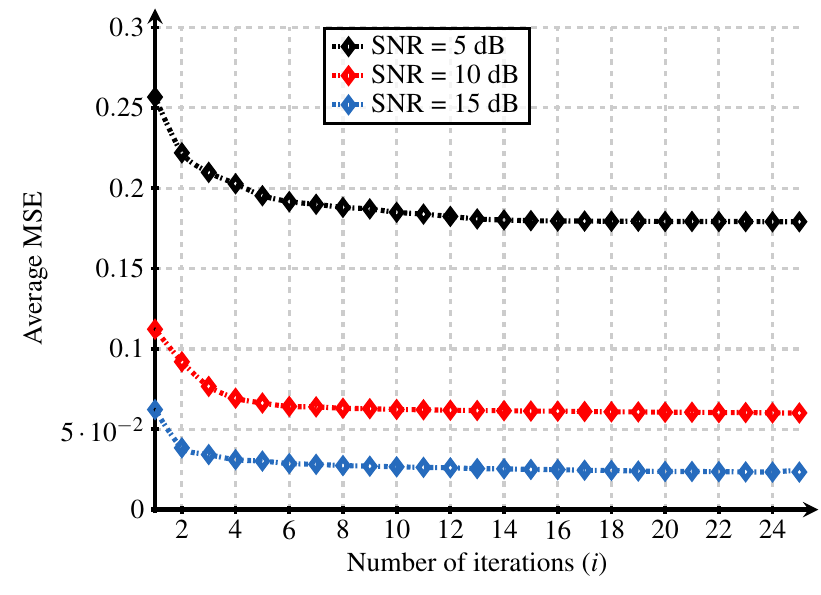} 
\vspace{-4.4mm}
 \caption{Average MSE vs number of iterations $(i)$ for $K=3$ users, $N_{\text{r}}=4$, $N_{\text{t}}=9$, $N=25$, $\text{SNR}\;(\text{dB})\in{\lbrace 5, 10, 15\rbrace}$, $L = 32$ imperfect CSI, and $N_{\text{s},k}[\ell]=2,\forall k,\ell$. }
  \label{fig4}
\end{figure}

In the last experiment, we empirically evaluate the convergence of \Cref{PG}. Specifically, \Cref{fig4} illustrates the evolution of the average \ac{MSE} (sum \ac{MSE} divided by $K=3$ users) concerning the number of iterations performed in the alternating procedure. We considered three different \ac{SNR} levels, namely $\text{SNR}\;(\text{dB}) \in  {\lbrace 5, 10, 15\rbrace}$.  Note that convergence is achieved after $25$ iterations in all cases.
As mentioned in \Cref{Section_conv}, proving global optimality for the algorithm is challenging. However, in our simulations, we have observed that the algorithm converges  rapidly within the first few iterations.

The conducted simulations reveal several remarkable results, which are summarized below:

\begin{enumerate}
\item The proposed PG-based approach outperforms the random configuration of the IRS-phase shift matrix, approaching the system capacity (\Cref{fig3a}).

\item \textit{The key is in the phase changes}. The proper configuration of our passive IRS does not lead to a big gap w.r.t. the approach where phases and amplitudes are properly changed in an active (power-limited) IRS, in the strategy termed as \ac{AF O-Ps} (\Cref{fig3a}).
\item  \textit{The placement of the IRS is crucial}.  The gains provided by the IRS become lower when
the IRS-user links become worse, i.e., the enhancement provided by the IRS is conditioned to the rank of the  IRS-user channels (\Cref{tablagananciaIRS}).
\item The use of low bit-resolution discrete phase shifts at the IRS is effective in achieving a substantial system sum rate that is comparable to the ideal scenario with continuous phase shifts (\Cref{fig3ar} and \Cref{figbits}).

    \item The performance gain achieved by the proposed solution over an IRS with random phase shifts saturates with a large number of IRS elements, depending on the wideband ($L$ subcarriers) $K$-users-MIMO setup (\Cref{fig3c}).

\item The proposed solution in Algorithm 1 is robust for imperfect CSI as it incorporates the statistics of channel estimation errors and achieves a substantial gain over a non-robust strategy (\Cref{fig4a} and \Cref{fig4b}). Additionally, the developed approach is independent to the modeling of the statistics of the errors.

\end{enumerate}

\section{Conclusions}\label{Section VII}
The design of an \ac{IRS}-aided wideband \ac{MU} \ac{MIMO} system under imperfect \ac{CSI} has been addressed in this paper. 
An innovative alternating minimization algorithm has been proposed to configure the frequency-flat \ac{IRS} phase-shift matrix as well as the wideband \ac{BS} precoders and user filters. The algorithm minimizes the \ac{MSE} between the symbols sent by the users and those received at the \ac{BS} in the dual \ac{MAC}. Imperfect \ac{CSI} is assumed and the available information on \ac{CSI} errors statistics is incorporated into the system design. The results show reasonable gains in terms of both the achievable rate and the \ac{MSE} over baseline strategies. Specifically, the deployment of the \ac{IRS} and the adequate configuration of the phase shift matrix provides significant performance gains 
with respect to non-\ac{IRS} conventional systems with \ac{MRT} precoding.

\section*{Acknowledgments}

This work has been supported by grants PID2019-104958RB-C42 (ADELE),  PID2022-137099NB-C42 (MADDIE), and BES-2017-081955 funded by MCIN/AEI/10.13039/501100011033. José P. González-Coma thanks the Defense University Center at the Spanish Naval Academy for all the support provided for this research.

\bibliographystyle{IEEEtran}
\bibliography{bmc_article}

\newpage

\vspace{11pt}

\vspace{11pt}

\vfill

\end{document}